\documentclass[longauth]{aa} 

\usepackage[utf8]{inputenc}
\usepackage[T1]{fontenc}

\usepackage{graphicx}
\usepackage{txfonts}
\usepackage{hyperref}
\hypersetup{unicode}
\hypersetup{breaklinks=true}
\hypersetup{colorlinks,					
	linkcolor={red!50!black},
	citecolor={blue!50!black},
	urlcolor={blue!80!black}}

\usepackage{lipsum}                     
\usepackage{xargs}                      
\usepackage[pdftex,dvipsnames]{xcolor}  

\usepackage{fdsymbol}
\usepackage{booktabs}
\usepackage{placeins}
\usepackage{siunitx}
\usepackage{float}
\usepackage[none]{hyphenat} 
\usepackage[nameinlink]{cleveref}
\usepackage{pdfpages}

\usepackage{dblfloatfix}
\usepackage{orcidlink}

\begin{document}

\newcommand{\obliquitybrown}[0]{$\lambda=2.8\pm3.1$ deg}
\newcommand{\Na}{Na\,{\textsc{i}}}
\newcommand{\Ha}{H\,{\textsc{$\alpha$}}}

\newcommand{\obliquityresult}[0]{$\lambda =  -0.09^{+0.31}_{-0.32}$ deg}
\newcommand{\vsiniresult}[0]{$v\sin(i_{\star}) = 7.42^{+0.11}_{-0.11} $ km / s}

   \title{Precise measurement of WASP-31 b's Rossiter-McLaughlin effect and characterization of the planet transmission spectra\thanks{Based on guaranteed time observations collected at the European Southern Observatory under ESO programs 108.2254.003 \& 110.24CD.003 by the ESPRESSO Consortium.}}
    \titlerunning{Precise measurement of WASP-31 b's transmission spectra}

   \author{M. Steiner\thanks{\href{mailto:Michal.Steiner@unige.ch}{Michal.Steiner@unige.ch}}\inst{\ref{inst:geneva}} \orcidlink{0000-0003-3036-3585} \and
          V. Bourrier\inst{\ref{inst:geneva}}\orcidlink{0000-0002-9148-034X} \and
          D. Ehrenreich\inst{\ref{inst:geneva}}\orcidlink{0000-0001-9704-5405} \and
          W. Dethier\inst{\ref{inst:porto}, }\orcidlink{0009-0007-3622-0928} \and
          H. Chakraborty\inst{\ref{inst:geneva}}\orcidlink{0000-0002-5177-1898} \and
          S. Pelletier\inst{\ref{inst:geneva}}\orcidlink{0000-0002-8573-805X} \and %
          M. Lendl\inst{\ref{inst:geneva}}\orcidlink{0000-0001-9699-1459} \and
          B. Akinsanmi\inst{\ref{inst:geneva}}\orcidlink{0000-0001-6519-1598} \and
          R. Allart\inst{\ref{inst:geneva},\ref{inst:montreal}}\orcidlink{0000-0002-1199-9759}\thanks{SNSF Postdoctoral Fellow} \and
          J. M.~Almenara\inst{\ref{inst:geneva}}\orcidlink{0000-0003-3208-9815} \and
          S. Cristiani\inst{\ref{inst:inaf}}\orcidlink{0000-0002-2115-5234} \and   
          J. I. Gonz\'alez Hern\'andez\inst{\ref{inst:iac},\ref{inst:ull}}\orcidlink{0000-0002-0264-7356} \and
          P. D. Marcantonio\inst{\ref{inst:inaf}}\orcidlink{0000-0003-3168-2289} \and
          C. J. A. P. Martins\inst{\ref{inst:porto},\ref{inst:porto_center_astrophysics}} \and
          L. Mishra\inst{\ref{inst:zalando}}\orcidlink{0000-0002-1256-7261} \and
          D. Mounzer\inst{\ref{inst:geneva}}\orcidlink{0000-0002-8070-2058} \and 
          M. R. Zapatero Osorio\inst{\ref{inst:madrid}}\orcidlink{0000-0001-5664-2852} \and
          E. Palle\inst{\ref{inst:iac},\ref{inst:ull}}\orcidlink{0000-0003-0987-1593} \and
          F. Pepe\inst{\ref{inst:geneva}}\orcidlink{0000-0002-9815-773X} \and
          A. Psaridi\inst{\ref{inst:geneva},\ref{inst:ice},\ref{inst:ieec}}\orcidlink{0000-0002-4797-2419} \and
          N. C. Santos\inst{\ref{inst:porto},\ref{inst:porto_astronomydepartment}}\orcidlink{0000-0003-4422-2919} \and 
          J. V. Seidel\inst{\ref{inst:nice}, \ref{inst:eso_chile}}\orcidlink{0000-0002-7990-9596}\thanks{ESO Fellow} \and
          A. Sozzetti\inst{\ref{ins:oato}}\orcidlink{0000-0002-7504-365X} \and
          V. Vaulato\inst{\ref{inst:geneva}}\orcidlink{0000-0001-7329-3471} \and
          G. Viviani\inst{\ref{inst:epfl}}\orcidlink{0009-0001-6201-2897} \and 
          J. Yu\inst{\ref{inst:epfl}}\orcidlink{0009-0001-7217-8006}
          }

   \institute{
   Observatoire de l'Universit\'{e} de Gen\`{e}ve, Chemin Pegasi 51, 1290 Versoix, Switzerland\label{inst:geneva}
   \and
   Instituto de Astrof\'{\i}sica e Ci\^{e}ncias do Espa\c{c}o, Universidade do Porto, Rua das Estrelas, 4150-762 Porto, Portugal\label{inst:porto}
   \and 
    D\'{e}partement de Physique, Institut Trottier de Recherche sur les Exoplan\`{e}tes, Universit\'{e} de Montr\'{e}al, Montr\'{e}al, Qu\'{e}bec, H3T 1J4, Canada\label{inst:montreal}
    \and
    INAF-Astronomical Observatory, via Tiepolo 11, I-34143 Trieste, Italy\label{inst:inaf}
    \and
    Instituto de Astrof\'{\i}sica de Canarias (IAC), 38205 La Laguna, Tenerife, Spain \label{inst:iac}
    \and
    Departamento de Astrof\'{\i}sica, Universidad de La Laguna (ULL), 38206 La Laguna, Tenerife, Spain \label{inst:ull}
       \and
    Centro de Astrof\'{\i}sica da Universidade do Porto, Rua das Estrelas, 4150-762 Porto, Portugal\label{inst:porto_center_astrophysics}
    \and
   Zalando Switzerland, Hardstrasse 201, 8005 Z\"{u}rich\label{inst:zalando} 
   \and
    Centro de Astrobiolog\'{\i}a, CSIC-INTA
    Camino Bajo del Castillo, s/n
    E-28692 Villanueva de la Ca\~{n}ada, Madrid\label{inst:madrid}
    \and
    Institute of Space Sciences (ICE, CSIC), Carrer de Can Magrans S/N, Campus UAB, Cerdanyola del Valles, E-08193, Spain\label{inst:ice}
    \and
    Institut d'Estudis Espacials de Catalunya (IEEC), 08860 Castelldefels (Barcelona), Spain\label{inst:ieec}
    \and
    Departamento de F\'{\i}sica e Astronomia, Faculdade de Ci\^{e}ncias,
Universidade do Porto, Rua do Campo Alegre, 4169-007 Porto,
Portugal \label{inst:porto_astronomydepartment}
    \and
    Laboratoire Lagrange, Observatoire de la C\^{o}te d'Azur, CNRS, Universit\'{e} C\^{o}te d'Azur, Nice, France\label{inst:nice}
    \and
    European Southern Observatory, Alonso de C\'{o}rdova 3107, Vitacura, Regi\'{o}n Metropolitana, Chile\label{inst:eso_chile}
    \and 
    INAF -- Osservatorio Astrofisico di Torino, via Osservatorio 20, I-10025, Pino Torinese, Italy \label{ins:oato}
    \and
    Institute of Physics, \'{E}cole Polytechnique Federale de Lausanne (EPFL), Observatoire de Sauverny, Chemin Pegasi 51b, 1290 Versoix, Switzerland\label{inst:epfl} 
    \\
    }

   \date{Received ...; accepted ...}

 
  \abstract
   {Hot Jupiters are ideal natural laboratories to investigate atmospheric composition and dynamics. However, high-resolution transmission spectroscopy is currently limited by our capability of removing planet-occulted line-distortion (POLD) contamination from the signal.}
   {In this paper, we aim to characterize the transmission spectrum of WASP-31 b from two and a half transits observed with the ESPRESSO spectrograph at the VLT.}
   {The Rossiter-McLaughlin (RM) signature was analyzed using the RM “revolutions” method. Before extracting the transmission spectrum of the planet, we corrected the dataset for telluric lines using \texttt{molecfit} and further modeled the POLD deformations using \texttt{EvE}.}
   {We confirm the planet low sky-projected spin-orbit angle from previous studies and further refine its value to \obliquityresult. We do not detect any species (including previously detected species such as K or CrH) in the planetary atmosphere. In most cases the non-detections are due to the strong POLDs contamination or lack of observable lines in the ESPRESSO wavelength range, and so previous detections cannot be ruled out.}
   {Planet-occulted line-distortion contamination continues to be the main limitation of high-resolution transmission spectroscopy for species present in both the star and the planet, hindering atmospheric detections even with state-of-the-art models, in particular for planets with a low sky-projected spin-orbit angle. Developing advanced techniques to isolate planetary signatures is of utmost importance in the advent of ELT-like observations.}


   \maketitle
%

\section{Introduction}
Hot Jupiters have become one of the most studied types of planet, as they are a great benchmark for planetary-formation and -evolution models, atmospheric characterization, and even atmospheric dynamics. Early studies show that hot Jupiters are perfect targets for atmospheric characterization studies, owing to their large atmospheric scale height. One of the most prominent methods for atmosphere detection is to use transit observations \citep{winn2014} to extract the transmission spectra.

Transmission spectroscopy was first proposed as a means for probing atmospheres of exoplanets by \cite{seager2000,brown2001}. This technique was used first by \citet{charbonneau2002} on HD 209458 b, a hot Jupiter, finding a sodium signature in the atmosphere in this planet, and is now the most prominent method for atmosphere detections around exoplanets. Following this discovery, Subaru/HDS was used to confirm this feature \citep{snellen2008}. Further analysis of HD 209458 b by high-resolution fiber-fed spectrographs such as HARPS \citep{casasayas-barris2020} or ESPRESSO \citep{casasayas-barris2021}, however, put the detection by \cite{charbonneau2002} in doubt.

The high-resolution transmission spectra were contaminated by planet-occulted line distortions \citep[POLDs,][]{dethier2023} resulting from the absorption by the planetary opaque layers of locally nonhomogeneous stellar spectra. When computing transmission spectra, the discrepancies between the occulted local stellar spectra and the disk-integrated stellar spectrum induce large distortions, which, without proper modeling, can lead to biased atmospheric property derivations \citep{dethier2023}. As shown by \cite{carteret2024}, in the case of HD 209458 b the high-resolution data can be reconciled with the medium-resolution sodium detection \citep{charbonneau2002, sing2008, vidal-madjar2011, santos2020, morello2022}, with the latter likely arising from the wings of the sodium lines and the former arising from the POLD in the core of the line.

To properly account for POLDs, models typically include two significant stellar line-distortion effects. First is the stellar rotation \citep[Rossiter-McLaughlin (RM)][]{rossiter1924, mclaughlin1924} effect, which affects the local and disc-integrated stellar spectra differently. During transit, the planet partially blocks either mainly blueshifted or redshifted local spectra, while the disc-integrated spectrum is not shifted, but broadened. The second stellar line-distortion effect is the variation of the line profiles over the apparent stellar surface. It affects the spectral lines’ shape locally due to the line-of-sight scanning varying stellar atmosphere depths as a function of the limb angle. This effect is typically weaker than the effect of rotation \citep[e.g.,][]{casasayas-barris2021}, but it still needs to be properly accounted for. Accounting for POLDs is thus a necessary step to detect atmosphere in transmission spectra contaminated by them and to further characterize the atmospheric dynamics through the line-profile shape \citep{seidel2021, seidel2023, seidel2025}. Typically, this means modeling the deformation \citep[e.g.,][]{casasayas-barris2017,chen2020}, estimating its effect \citep{mounzer2022}, or masking the POLD affected regions completely (thus also removing planetary signal). However, correcting the transmission spectra for these distortions through POLDs modeling still leaves altered line profiles, which can be reconciled by modeling the atmosphere and POLD distortions at the same time \citep{dethier2024}.

To advance progress within the field of high-resolution transmission spectroscopy, the ESPRESSO guaranteed time observation (GTO) consortium dedicated significant time to transit high-resolution observations. As a standard approach, each of the selected targets has been followed for two transits using the VLT/ESPRESSO \citep{pepe2010,pepe2014}. Among these results, we can name detections of many species in the atmosphere of WASP-76 b \citep{tabernero2021}, the detection of time-asymmetry in atmospheric signatures for the first time \citep{ehrenreich2020}, and an analysis of the atmospheric dynamics in the same planet by \cite{seidel2021}. \cite{borsa2021} succeeded in using the 4-UT mode of ESPRESSO on the commissioning data of WASP-121 b, which later led to the detection of a high-velocity jet in the egress data by \cite{seidel2023,seidel2025}. \cite{azevedosilva2022} detected barium in both WASP-76 b and WASP-121 b transmission spectra, the heaviest element found in the planet's atmosphere to date. Regarding the RM effect, \cite{bourrier2022} validated the measurement of the polar orbit of a warm Neptune orbiting an M dwarf, GJ 436 b, while \cite{allart2020} and \cite{cristo2022} determined the retrograde orbit of WASP-127 b. Finally, the aforementioned \cite{casasayas-barris2021} showed the significant effect POLDs have on the high-resolution transmission spectra.

\begin{figure}
    \centering
    \includegraphics[width=\columnwidth]{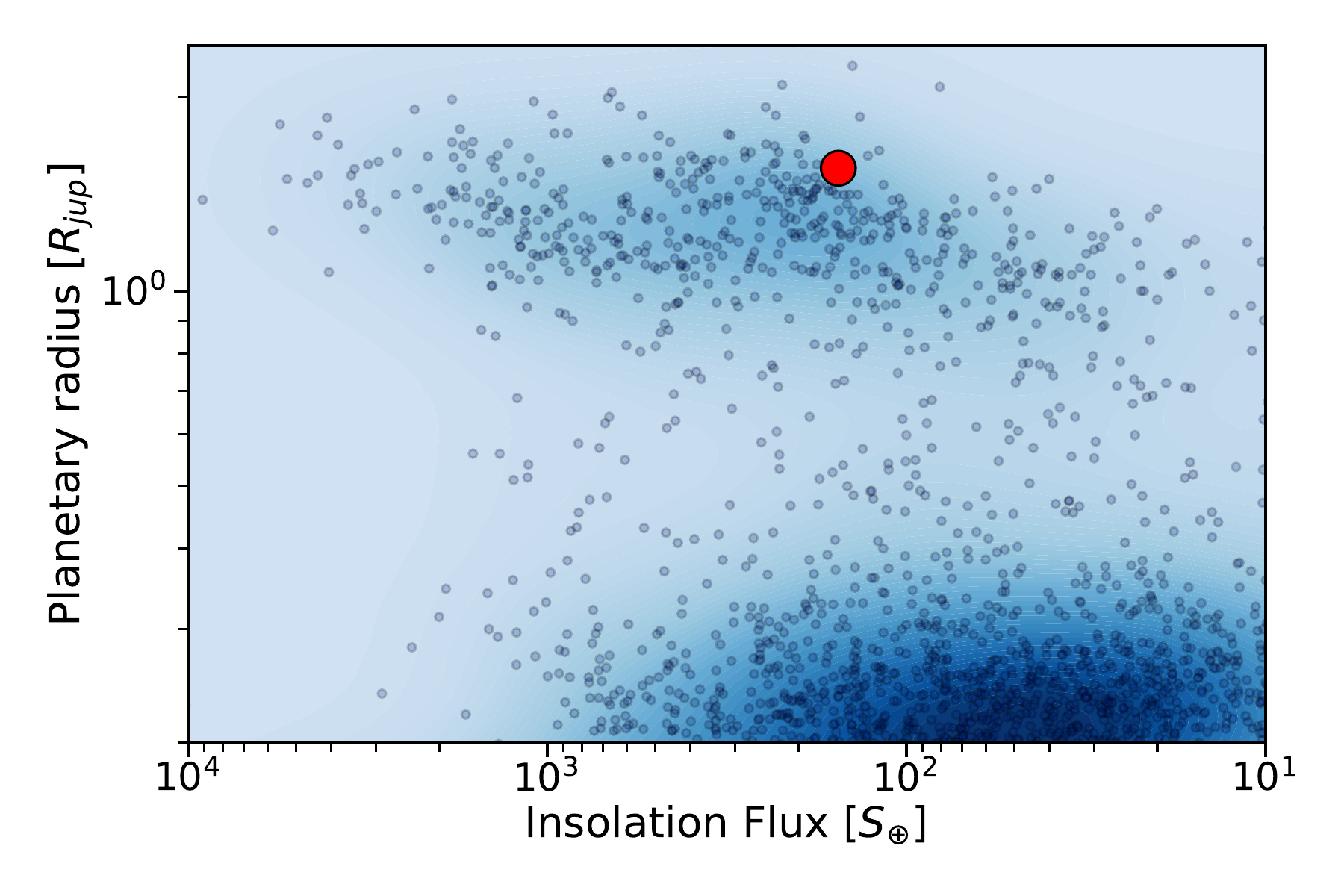}
    \caption{Location of WASP-31b (red point) in radius versus insolation flux diagram of known exoplanets (small black points), zoomed-in on the hot-Jupiter region. Extracted from NASA Exoplanet archive on 24 March, 2025}.
    \label{fig:radius_insolation_flux}
\end{figure}

In this paper, we present the first analysis of the ESPRESSO dataset of WASP-31b (\autoref{fig:radius_insolation_flux}; \autoref{tab:system_parameters}), an inflated hot Jupiter, as part of the ESPRESSO GTO program. This planet has already been observed with multiple instruments with contradicting results. First, \cite{brown2012} measured the projected spin-orbit alignment of WASP-31 b to \obliquitybrown. In the planet's atmosphere, \cite{sing2016} detected a strong feature of potassium using HST/STIS, which was then revisited by \cite{gibson2017} and \cite{gibson2019}, which were not able to confirm it with VLT/FORS2 (former) and VLT/UVES (latter). Furthermore, \cite{braam2021} suggested the potential detection of chromium hydride (CrH) in this planet, which was confirmed by \cite{flagg2023}. This is the first time CrH has been detected in a planet's atmosphere, with only a second tentative detection of CrH by \cite{jiang2024} in HAT-P-41 b. 

\begin{table*}
        \begin{center}
        \caption{System parameters}
        \resizebox{2\columnwidth}{!}{
\begin{tabular}{llll}
        \toprule
        Stellar Parameter & This paper (\autoref{subsec:photometry}) & Reference value & Reference \\
        \midrule
        Temperature $(T_\mathrm{eff})$ & --- & 6302.0 $\pm 102.0$ K &  \cite{anderson2011} \\
        Radius $(R_\mathrm{s})$& $1.252^{+0.030}_{-0.030} R_{\odot}$ &1.252 $\pm 0.033$ $R_{\odot}$ &  \cite{anderson2011} \\         
        Mass $(M_\mathrm{s})$& $1.240_{-0.091}^{+0.097}M_{\odot}$ & 1.163 $\pm 0.026$ $M_{\odot}$ &  \cite{anderson2011} \\ %
        Age & --- &1.00 $^{+3.00}_{-0.50}$ Gyr & \cite{bonomo2017} \\
        Systemic velocity $(\gamma)$ & --- & -0.1 $\pm 1.1$ km / s & \cite{gaiadrs3} \\
        Distance to the system $(d)$ & --- & 383.8 $\pm 6.4$ pc & \cite{stassun2019} \\
        \toprule
        Planetary parameter & This paper (\autoref{subsec:photometry}) & Reference value & Reference \\
        \midrule
        Radius $(R_\mathrm{p})$ & $1.507_{-0.037}^{+0.036}$ $R_\text{jup}$ & 1.549 $\pm 0.050$ $R_\text{jup}$ &  \cite{anderson2011} \\  
        Mass $(M_\mathrm{p})$ & --- & 0.478 $\pm 0.029$ $M_\text{jup}$ &  \cite{anderson2011} \\
        Density $(\rho_\mathrm{p})$& --- & 0.172 $\pm 0.019$ g / cm3 &  \cite{anderson2011} \\
        Semimajor-axis $(a)$& $0.047\,9_{-0.001\,2}^{+0.001\,2}$ $ AU$ & 0.046\,59 $\pm 0.000\,35$ AU &  \cite{anderson2011} \\
        Eccentricity $(e)$& --- & 0.0 (unitless, fixed) &  \cite{anderson2011} \\
        Equilibrium temperature $(T_\mathrm{eq})$& --- & 1575 $\pm 32$ K & \cite{anderson2011} \\
        Keplerian RV semi-amplitude $(K)$  & --- & 59.4 $^{+2.8}_{-2.9}$ m / s & \cite{bonomo2017} \\
        \toprule
        Ephemeris parameter & This paper (\autoref{subsec:photometry}) & Reference value & Reference \\
        \midrule
        Transit center $(T_\mathrm{c})$& $2\,459\,606.719\,11_{-0.000\,11}^{+0.000\,14}$ d & 2\,457\,277.092\,26 $\pm 0.000\,12$ d & \cite{kokori2023} \\ 
        Planetary period $(P)$& $3.405\,887\,90_{-0.000\,000\,62}^{+0.000\,000\,63}$ d & 3.405\,887\,50 $\pm 0.000\,000\,27$ d & \cite{kokori2023} \\ 
        Transit length (partial, $T_\mathrm{14}$) & $2.651_{-0.012}^{+0.010}$ h & 2.647 $\pm 0.031$ h &  \cite{anderson2011} \\ 
        Transit depth $(\delta)$& $0.0153070_{-0.000110}^{+0.000096}$ (unitless) & 0.016\,15 $\pm 0.000\,27$ (unitless) &  \cite{anderson2011} \\ 
        \toprule
        RM anomaly analysis & This paper (\autoref{subsec:RM_anomaly}) & Reference value & Reference \\
        \midrule
        Projected spin-orbit angle $(\lambda)$& $0.13^{+0.33}_{-0.34}$ deg & 2.8 $\pm 3.1$ deg & \cite{brown2012} \\
        Projected rotation speed $(\text{v}_\text{eq}\sin(i_{\star}))$ & $6.90^{+0.10}_{-0.10}$ km / s & 7.5 $\pm 0.7$ km / s & \cite{brown2012} \\
        Systemic velocity 2022-01-26 & -0.2801 $\pm 0.0013$ km / s & -0.1 $\pm 1.1$ km / s & \cite{gaiadrs3} \\
        Systemic velocity 2022-02-12 & -0.2447 $\pm 0.0014$ km / s & -0.1 $\pm 1.1$ km / s & \cite{gaiadrs3} \\
        Systemic velocity 2023-02-25 & -0.2652 $\pm 0.0012$ km / s & -0.1 $\pm 1.1$ km / s & \cite{gaiadrs3} \\
        \bottomrule
\end{tabular}
                }
        \end{center}
    \begin{minipage}{\linewidth}
    \vspace{0.1cm}
    \small Notes: System parameters, as obtained by our analysis (second column, \autoref{subsec:photometry} for photometry and \autoref{subsec:RM_anomaly} for RM analysis), and the set we used in this work (third column; fourth column for the reference paper). Solutions for planet eccentricity were fixed at 0 where possible, though \cite{bonomo2017} reported an upper limit of 0.047. Eccentricity plays a role when shifting to the rest frame of the planet. However, the level is negligible even for small eccentricities. We used the set of stellar and planetary parameters following the literature (third column), while for the ephemeris we used our refined values (second column) after testing that they were compatible. For POLD simulation, we also used values obtained by our analysis due to the higher precision level.
    \end{minipage}
 
        \label{tab:system_parameters}
\end{table*}

We focused our analysis on the transmission spectrum of WASP-31 b, in particular the search for \Na\ and \Ha. Furthermore, we utilized the cross-correlation function (CCF) to search for iron (Fe) and CrH. This paper is organized as follows. In \autoref{sec:observation}, we present the analyzed dataset and the observational conditions. In \autoref{sec:methodology}, we describe the methodology used for data analysis, with simultaneous photometry analysis in \autoref{subsec:photometry}, Rossiter-McLaughlin analysis in \autoref{subsec:RM_anomaly}, and transmission spectroscopy extraction in \autoref{subsec:tsextraction}. The results of transmission spectroscopy are presented in \autoref{sec:transmission_spectroscopy} and are further discussed in \autoref{sec:discussion}. Finally, we provide concluding remarks in \autoref{sec:conclusion}.

\section{Observations}
\label{sec:observation}

\begin{table*}
	\begin{center}
    \caption{Observation log table}
 \resizebox{2\columnwidth}{!}{
		\begin{tabular}{llllrrrrr}
			\toprule
			Night & Night \# & Instrument & prog. ID & Exposure time [s] & In/Total \# & Seeing ["] & Airmass & S/N \\
			\midrule
			2022-01-26 & 1 & ESPRESSO & 108.2254.003 & 500 & 17/31 & 0.50 - 0.62 - 0.89 & 1.00 - 1.09 - 1.72 & 21 - 28 - 30 \\
			2022-02-12 & 2 & ESPRESSO & 108.2254.003 & 500 & 6/19 & 0.54 - 0.79 - 1.10 & 1.05 - 1.23 - 1.73 & 20 - 26 - 31 \\
			2023-02-25 & 3 & ESPRESSO & 110.24CD.003 & 500 & 17/31 & 0.44 - 0.66 - 1.06 & 1.00 - 1.06 - 1.60 & 20 - 28 - 31 \\
			\bottomrule
		\end{tabular}
        }
	\end{center}
    \begin{minipage}{\linewidth}
    \vspace{0.1cm}
    \small Notes: The columns from left to right show the date of the observing night, the number index of a given night, the instrument used, the program number, the exposure time, the number of in-transit and total spectra on a given night, the seeing, the airmass, and the average S/N. Note that for seeing airmass and average S/N, we provide the value in min-median-max format. S/N is the average over all orders for a given instrument. For ESPRESSO, this does not account for double-order structure of the spectra, meaning the total S/N is generally higher by a factor of $\sqrt{2}$ at a given wavelength.
    \end{minipage}
    
	\label{tab:observation_log}
\end{table*}

We observed a total of three nights with ESPRESSO (as part of the GTO Working Group 2 (WG2)). The summary of the nights, including their respective program IDs, is provided in \autoref{tab:observation_log}, with a full observation log per spectrum shown in \autoref{fig:observation_log}. 

The first and third nights are of comparable quality, with similar observing conditions and setup. The second night was stopped mid-transit due to degraded weather conditions, leaving only a few in-transit exposures to provide information about the planet. All three nights were observed using the Unit Telescope 1 (UT1) and STAR+SKY fiber reference mode (fiber B on sky background). We used the data-reduction software (DRS) to reduce the data (\texttt{ESPRESSO DRS 3.0.0}).

\begin{figure}
    \centering
    \includegraphics[width=\columnwidth]{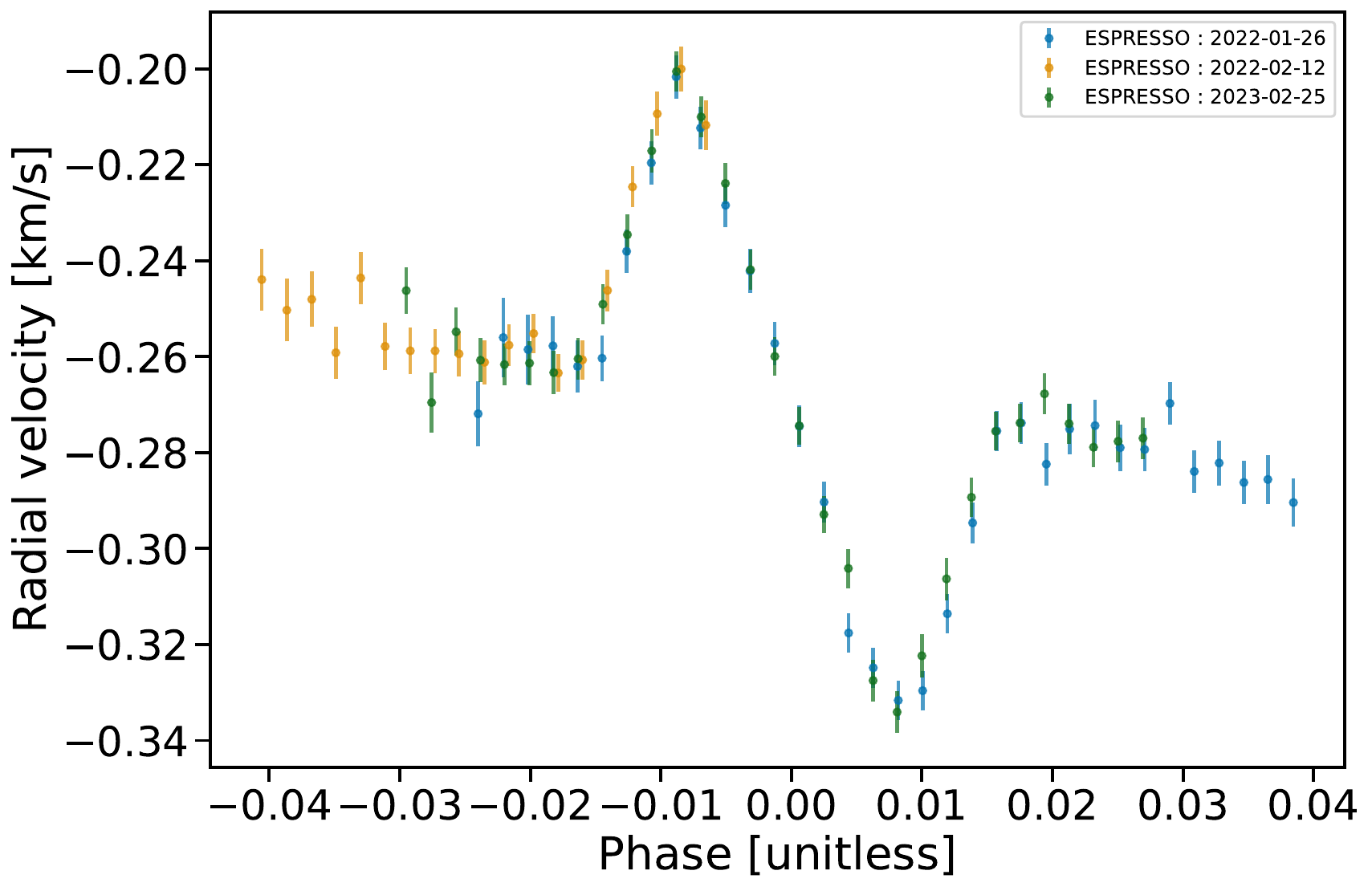}
    \caption{Radial-velocity observations of ESPRESSO data. The Rossiter-McLaughlin anomaly is visible, showing the typical signature of an aligned planet as previously deduced by \cite{brown2012} using a single HARPS transit. Nights are color-coded separately in blue (night \#1), orange (\#2), and green (\#3).}
    \label{fig:radial_velocity_plot}
\end{figure}

The ESPRESSO radial-velocity (RV) curve (\autoref{fig:radial_velocity_plot}) shows a clear Rossiter-McLaughlin effect, suggesting an aligned planet, as previously observed by \cite{brown2012}.  However, due to the larger mirror size, the precision of ESPRESSO RVs is much higher compared to those of the previous observation of HARPS. While two nights' of HARPS observations exist (the aforementioned data used by \cite{brown2012} and a single transit in the HEARTS survey; prog. ID: 097.C-1025(C), PI: Ehrenreich, D.), we did not use the HARPS data due to the lower S/N. In particular, the transmission spectra from HARPS are much more dominated by red noise due to the lower S/N, while the RM analysis does not provide further information.

Simultaneously to the ESPRESSO observations, we followed the target using the ECAM \citep{lendl2012} at the 1.2m Swiss telescope. While photometry analysis was not the focus of our study, these data were used to refine the ephemeris and to detect potential stellar activity during the transit.

\section{Methodology}
\label{sec:methodology}

In \autoref{subsec:photometry}, we discuss the simultaneous photometry analysis obtained with ECAM. Our methodology for transmission spectroscopy follows \cite{steiner2023}, which is based on the same pipeline\footnote{Available at \url{https://github.com/MichalSteiner/rats}, the pipeline is completely open-source for the purpose of transmission spectroscopy and RM anomaly analysis.}. However, multiple improvements to the pipeline have been made since then, which we describe below. Furthermore, RM effect analysis has been implemented in this pipeline recently (following \cite{cegla2016,bourrier2021}), so we describe the methodology used in this paper in \autoref{subsec:RM_anomaly}. 

\subsection{Simultaneous photometry analysis}
\label{subsec:photometry}
Simultaneously with the ESPRESSO observations, we obtained transit light curves using the ECAM \citep{lendl2012} at the ESO La Silla 1.2 m Swiss Telescope. These data were used in two ways during the analysis. First, each of the transit light curves was visually checked for potential stellar spots. Second, using the ECAM light curves combined with TESS, we attempted to refine the system parameters using CONAN \citep{lendl2017}. Our results for stellar, planetary, and ephemeris parameters (\autoref{tab:system_parameters}, second column) are in good agreement with previous observations (third column). Because our analysis does not include the RV data, we do not obtain the mass of the planet, which is important for velocity calculation when shifting rest frames. Our ephemeris parameters are compatible with \cite{kokori2023}. However, our work is incompatible with previous observations by \cite{patel2022}. We discuss the ephemeris choice used for our analysis in more detail in \autoref{subsec:ephemeric-choice}.

\begin{figure}
    \centering
    \includegraphics[width=\linewidth]{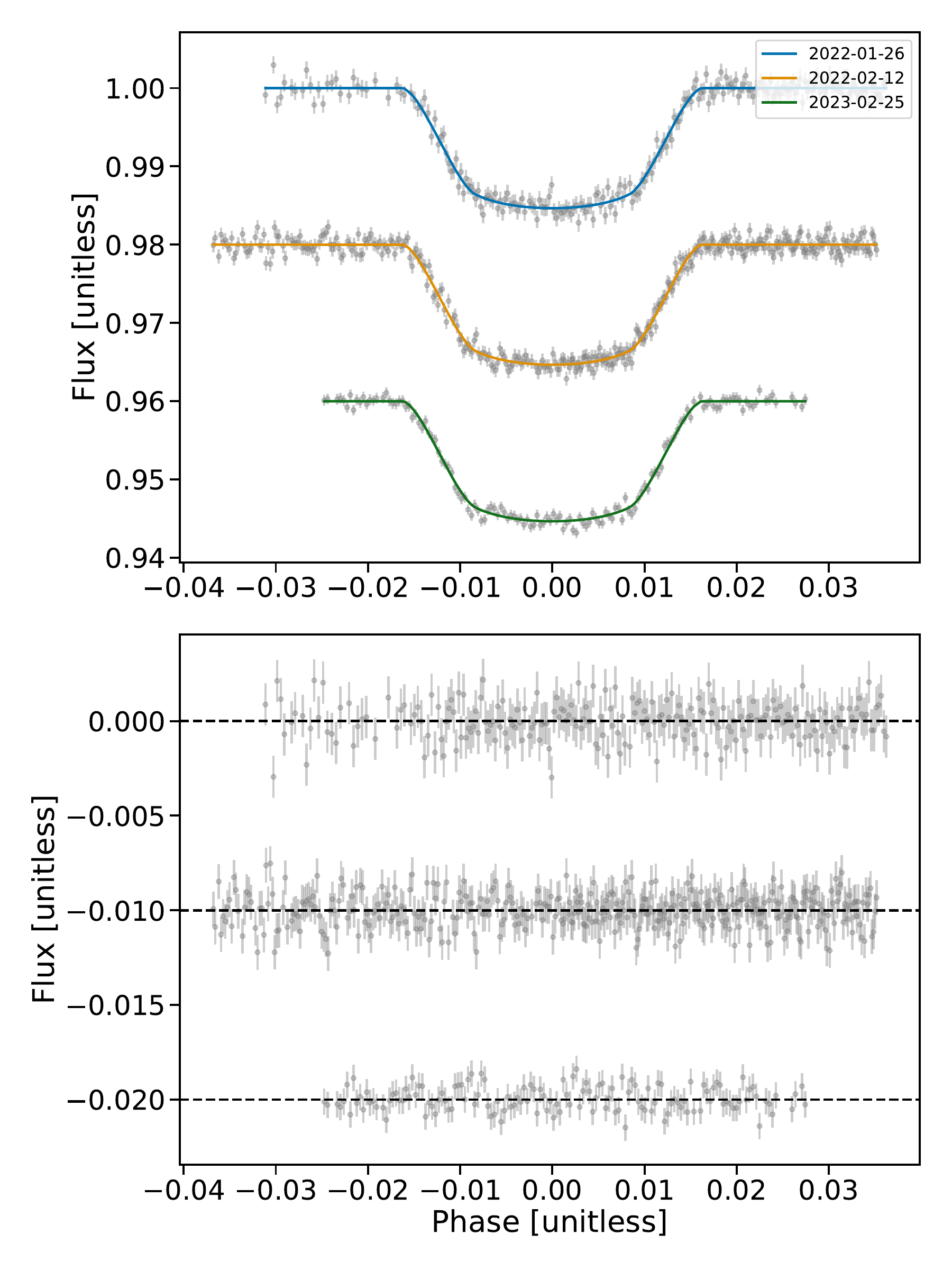}
    \caption{Observed simultaneous photometric light curves (top panels) from ECAM for the three ESPRESSO nights. The best-fit models from CONAN for each night are color-coded separately in blue (night \#1), orange (\#2), and green (\#3). The residuals for each night are shown in the bottom panel with an offset. The horizontal line is the continuum of 0.}
    \label{fig:light-curve}
\end{figure}

Thus, in our analysis, we used the stellar and planetary parameters provided in the third column of \autoref{tab:system_parameters}, particularly those by \cite{anderson2011}, which we combine with the ephemeris and RM anomaly values from our analysis. The resulting light curves and the best-fit model are shown in \autoref{fig:light-curve}.

\subsection{Analysis of RM anomaly}
\label{subsec:RM_anomaly}
To analyze the RM anomaly, we used the “revolutions” method \citep{bourrier2021}, which further builds on the “reloaded” method by \cite{cegla2016}. The RM anomaly analysis we performed was done on the \texttt{CCF\_SKYSUB\_A} products. These products are the output of the cross-correlation of the stellar template with the observed spectrum corrected for sky background and, as such, consist of an array of velocities, flux, and uncertainties. These CCFs are roughly Gaussian in shape, which, when fit, provides the observed RV of a given spectrum. We only used the combined CCF of all spectral orders in our analysis.

Before the analysis, we defined the light-curve model using \texttt{batman} \citep{kreidberg2015} with the parameters as reported in \autoref{tab:system_parameters}. For limb-darkening coefficients, we used \texttt{LDCU}\footnote{Available at \href{https://github.com/delinea/LDCU}{https://github.com/delinea/LDCU}} \citep{deline2022} to calculate the quadratic law coefficients, which were then passed to \texttt{batman}. We simulated a transit light curve for BJD timestamps of all CCFs.

The first step in our analysis is to normalize the CCFs to 1. To accomplish this, we fit a Gaussian profile with a constant term:
$G(A, \sigma, v_0) + C$,
where G is the Gaussian profile with amplitude (contrast) A, standard deviation $\sigma$ and mean value of $v_0$, while the term C is our continuum trend (assumed constant). By dividing by C, we normalized each CCF to 1. 

We furthermore scaled each CCF to their respective light-curve fluxes ($\Delta(t) = 1- \delta(t)$), as calculated before by \texttt{batman}. Here, $\Delta(t)$ is the normalized light-curve flux, and $\delta$ is the transit depth due to the planet at a given time (0 when the planet is not transiting). Thus, the final normalized CCF is given by
$$\text{CCF}_{i}(t) = \frac{\text{CCF}_\text{i; raw}}{C_i} \Delta_{i}(t)$$.

As the CCFs are in the barycenter of the Solar System, we needed to shift by the systemic velocity and the stellar velocity induced by the planet to the stellar rest frame. We calculated the systemic velocity for each night by fitting a master-out CCF\footnote{Calculated as unweighted average out-of-transit spectrum.}, following \cite{bourrier2021}, as there might be offsets between nights. All CCFs are first shifted by the stellar velocity, assuming a circular orbit given parameters from \autoref{tab:system_parameters}. Then we calculated the master-out CCF (average out-of-transit CCF), which we fit with a Gaussian profile. The mean value of this profile provides the systemic velocity of each night, which we used to shift our CCFs accordingly.

Then, we took the original set of CCFs in the solar barycentric rest frame, where we shifted by the previously calculated systemic velocity and stellar velocity induced by the planet. There, we calculated the master-out CCF ($M_\text{CCF; out}$). We then subtracted each CCF from the calculated master-out to obtain the residual CCF ($\text{CCF}_\text{res}$) and the intrinsic CCF ($\text{CCF}_\text{intr}$), which are scaled by transit depth. As $\text{CCF}_\text{intr}$ depends on $\delta_{i}$, we did not calculate $\text{CCF}_\text{intr}$ for out-of-transit CCFs.

$$\text{CCF}_\text{i,res} = (M_\text{CCF; out} - \text{CCF}_{i}(t))$$

$$\text{CCF}_\text{i,intr, in} = \frac{(M_\text{CCF; out} - \text{CCF}_{i}(t))}{\delta_{i}(t)}$$
  
To inspect the quality of each intrinsic in-transit CCF, we followed \cite{bourrier2021} with MCMC fitting of a Gaussian profile to the CCF. We used the \texttt{PYMC} package. We used non-informative priors, limited only by the range of the CCF output, which is possible to modify in the DRS. We used the No U-Turn Sampler (NUTS) in the MCMC, with 15000 burning steps \footnote{In \texttt{PYMC}, burning steps are called “tune” steps.} and 15000 drawing steps (meaning the first 50\% of steps are thrown away). We used 30 chains, which were run on each individual exposure. The final posterior plots are depicted in \autoref{fig:posteriors_intrinsicccf}. 

A visual inspection of the posterior plots (\autoref{fig:posteriors_intrinsicccf}) showed several poorly defined exposures, particularly the first and last exposures in the transit sequence. We removed these from the analysis. More information on this can be found in \autoref{subsec:pdf_intrinsic}. Furthermore, during the first night, an anomaly was found in the contrast shape of the raw $\text{CCF}_{i}$. This is likely connected to the higher asymmetry of the Gaussian profile during the first night, as opposed to the second and third nights, as the residuals show a clear systematic trend as well. We further discuss the contrast anomaly in \autoref{subsec:contrast-anomaly}.

After the removal of ill-defined $\text{CCF}_\text{i, intr}$, we proceeded with the global fitting of all in-transit $\text{CCF}_\text{i, intr}$. We fit for the contrast, FWHM, v$_\text{eq}\sin(i)$, and $\lambda$. We tested multiple-order polynomials of contrast and FWHM trends as a function of limb-angle $\mu$, following \cite{bourrier2021} to fit the $\text{CCF}_\text{i, intr}$. The tested models include a constant profile for all nights, night-specific constant profiles, a night-specific linear model, and quadratic models. Finally, as the second night holds only a few exposures, we also ran the linear and quadratic model assuming a zeroth order on the second night, specifically, to avoid overfitting. By comparing the BIC, we found the constant profile to be the preferred model. Therefore, we assumed the non-varying shape of the stellar CCF within the data, effectively neglecting the CLV effect, as based on our tests they can be neglected within the noise level. Thus, our final priors for the “revolutions” fit were:

\begin{center}
contrast $\sim$ \( \mathcal{U} \)$(0,1)$ \\,
FWHM $\sim$ \( \mathcal{U} \)$(0,15)$ kms$^{-1}$\\,
v$_\text{eq}\sin(i_\star)$ $\sim$ \( \mathcal{U} \)$(0,10)$ kms$^{-1}$\\,
$\lambda \sim$ \( \mathcal{U} \)$(-180,180)$ deg\\
\end{center},

where v$_{eq}\sin(i_\star)$ is the projected equatorial rotation velocity of the star and $\lambda$ is the projected spin-orbit angle. We again used non-informative priors on the contrast of the CCFs. We used an FWHM prior that is fully encompassing the FWHM measurement on the $M_\text{CCF; out}$ (12.5 kms$^{-1}$). As the local stellar velocity depends on both v$_{eq}\sin(i_\star)$ and $\lambda$, we used these as our jump parameters. The priors for these are again non-informative, although we could have used values provided by \cite{brown2017}. We did not use those priors due to the different methodology used to obtain these values, making our observations fully independent of each other. For v$_\text{eq}\sin(i_\star)$, we used a prior restricted only by the shape of the $M_\text{CCF; out}$. For $\lambda$, we used a uniform distribution across the full range of possible values. We assumed solid-body rotation. We used 30 chains for this MCMC analysis, with 15000 burning and drawing steps.

\begin{figure*}
    \centering
    \includegraphics[width=2\columnwidth]{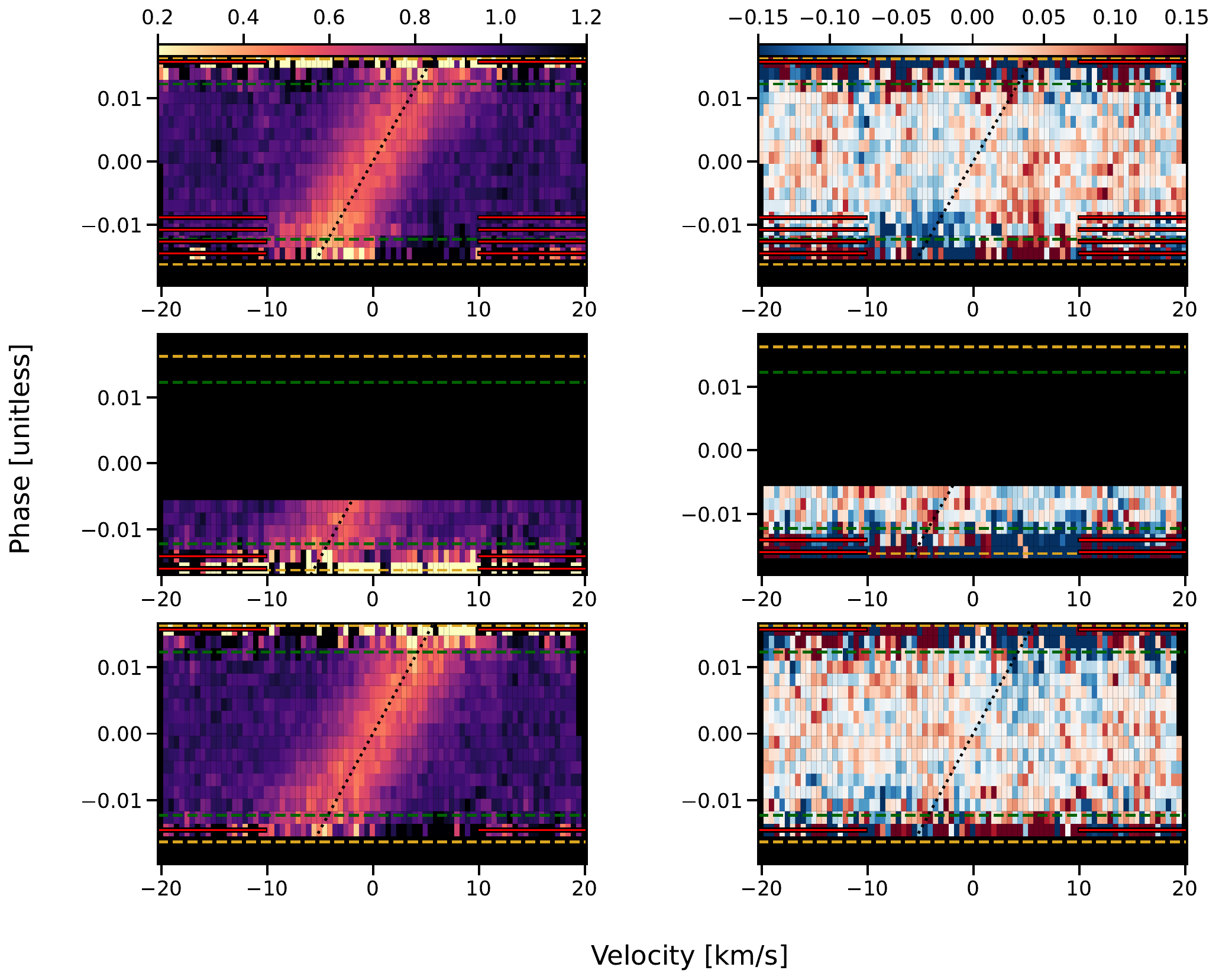}
    \caption{Intrinsic CCF per night (left panels) and their respective residuals (right panels). The RM signature is well visible during the transit as a bright track. The dashed horizontal lines correspond to T14 contact points (orange) and T23 (green). The solid red lines (with black outlines) correspond to spectra that were excluded from the analysis. The dotted black lines correspond to the expected local stellar velocity using the best-fit model. The first night's residuals show a systematic trend, likely due to line asymmetry.}
    \label{fig:intrinsic_CCF}
\end{figure*}

The resulting intrinsic CCFs and the residuals of subtracting the best-fit model obtained by the MCMC analysis from the intrinsic CCFs are shown in \autoref{fig:intrinsic_CCF}. For the second and third night, we see a good fit to the data, but the first night shows a systematic difference. This is likely due to the stellar lines being asymmetric during the first night. This could, for example, be due to a high spot coverage; however, a visual examination of the photometry (\autoref{fig:light-curve}) did not confirm this. We note that our method might not be sensitive enough during the ingress for spot detection.

\begin{figure}
    \centering
    \includegraphics[width=\columnwidth]{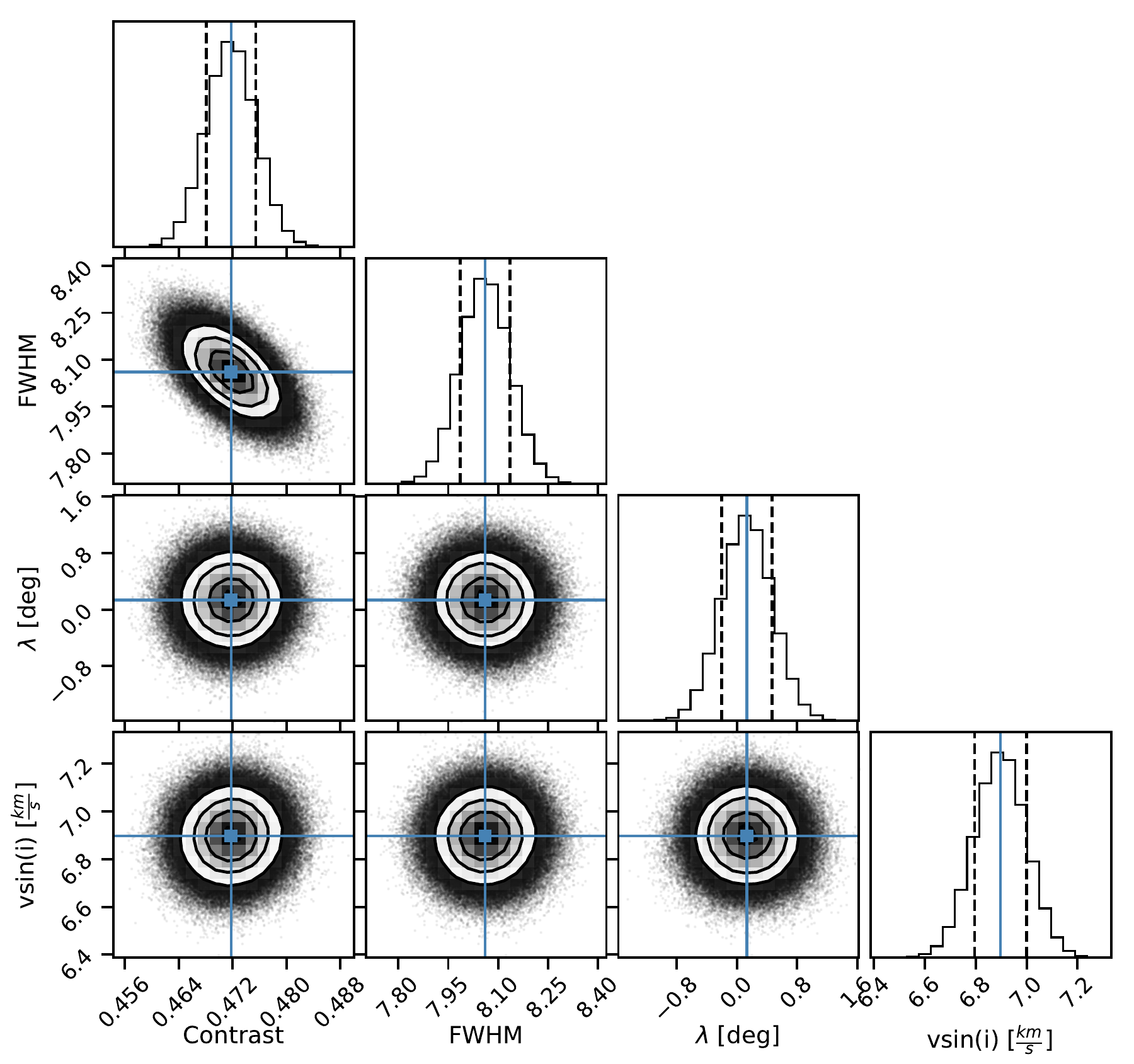}
    \caption{Final posterior corner plot of e revolutions analysis of RM anomaly. All values are well-defined, with an anticorrelation between the contrast and FWHM of the line profile. The contours show 1$\sigma$, a region containing 50\% of accepted steps and 2$\sigma$ levels, as defined by the \texttt{corner} package. The solid lines correspond to the median value of the posterior distribution function, which is used as the measured value.}
    \label{fig:corner_RMR}
\end{figure}

The resulting corner plot is provided in \autoref{fig:corner_RMR}. To obtain the resulting values for $\lambda$ and v$_{eq}\sin(i_\star)$, we used the median values of the posteriors. We used the highest density intervals (HDI) of 68.3\% to obtain the 1$\sigma$ values of each parameter. The results are summarized in \autoref{tab:RMR_results}.

\begin{table}[]
    \centering
    \caption{Results of RM analysis}
    \begin{tabular}{lrr}
    \toprule
    Variable & Value (This paper) & \cite{brown2012} \\
    \midrule
    $\lambda$ [deg]& $0.13^{+0.33}_{-0.34}$  & $2.8\pm3.1$ \\
    $v\sin(i_{\star})$ [km/s]& $6.90^{+0.10}_{-0.10}$ &  7.5 $\pm 0.7$ \\
    \bottomrule
    \end{tabular}
    \begin{minipage}{\linewidth}
    \vspace{0.1cm}
    \small Notes: Final values for projected obliquity $\lambda$ and v$\sin(i_\star)$, as analyzed with revolutions (second column), and its comparison with \cite{brown2012}. Our results are in good agreement, but more precise than those of \cite{brown2012}.
    \end{minipage}
    \label{tab:RMR_results}
\end{table}

Finally, we looked at the available TESS data to see whether the rotation period of the star could be constrained with a Lomb-Scargle periodogram analysis. Unfortunately, such attempts proved unsuccessful. Thus, the full 3D obliquity $\psi$ is left unknown. We note, however, that systems with $\lambda\approx0^\circ$ are rarely truly aligned \citep{attia2023}.

\subsection{Transmission-spectroscopy extraction}
\label{subsec:tsextraction}
\begin{figure}
    \centering
    \includegraphics[width=\columnwidth]{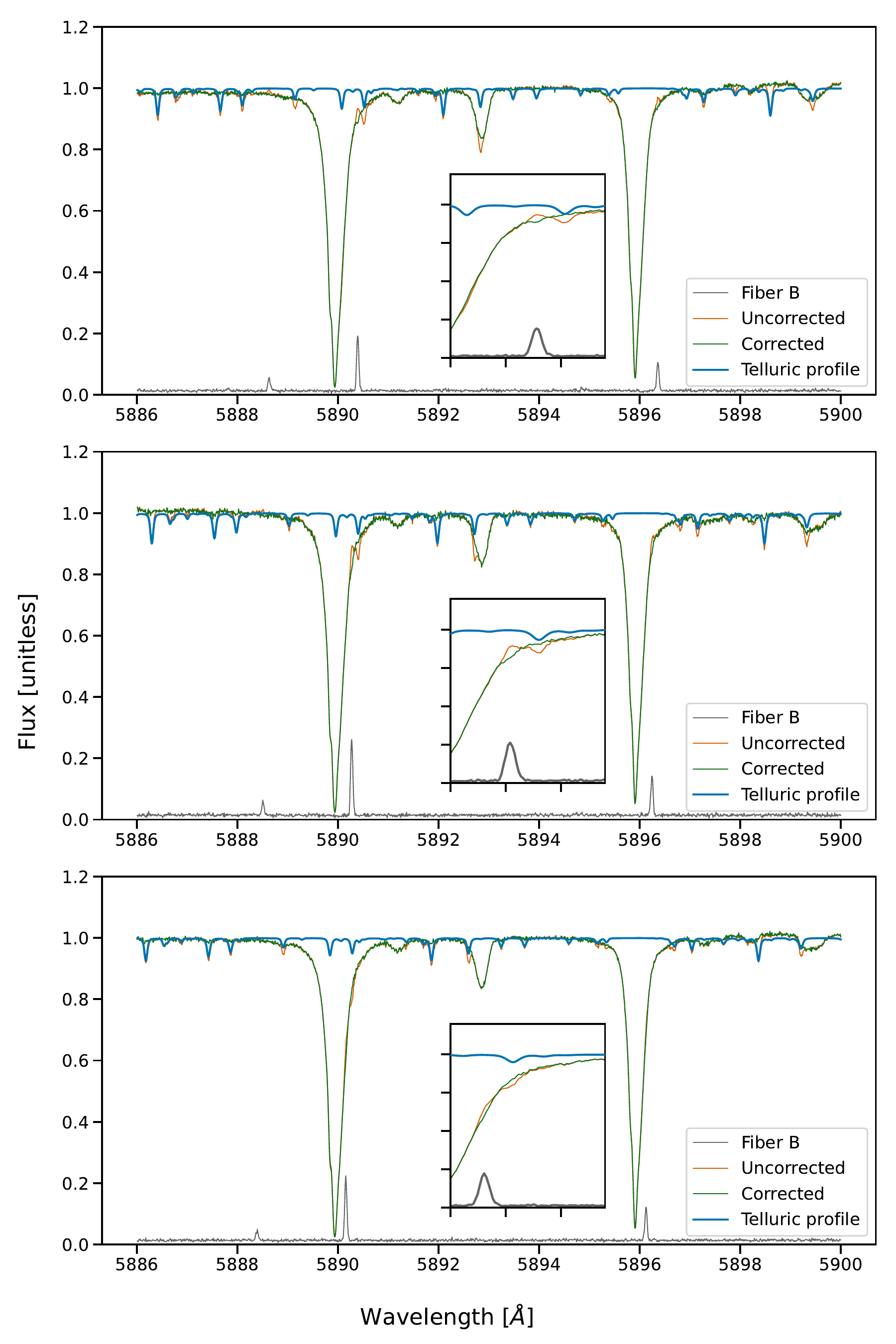}
    \caption{Correction for fiber B and telluric lines in the region of the sodium doublet. Each panel shows a respective night, with the master of raw spectra from fiber A (uncorrected for fiber B and telluric lines) in orange. In green, we show the master of fiber A spectra corrected for fiber B and telluric lines. The respective average telluric profile and fiber-B emission are shown in blue and gray, respectively. Please note that fiber B is on an arbitrary scale due to normalization. The zoomed-in image of the inset focuses on the wavelength region of 5890-5891 $\AA$}.
    \label{fig:earth_correction}
\end{figure}

The main change in our approach for extracting transmission spectroscopy is our use of the \texttt{S1D\_SKYSUB} products, that is, the stitched echelle spectrum that is subtracted by the fiber B spectrum considering the differing fiber throughput. This means that telluric emission features, such as those of sodium, should already be corrected, and no further correction is necessary. We confirmed this by calculating the average fiber B spectrum per night and visually checking the prominent emission lines against the average fiber A spectrum per night (\autoref{fig:earth_correction}). 

\begin{figure}
    \centering
    \includegraphics[width=\columnwidth]{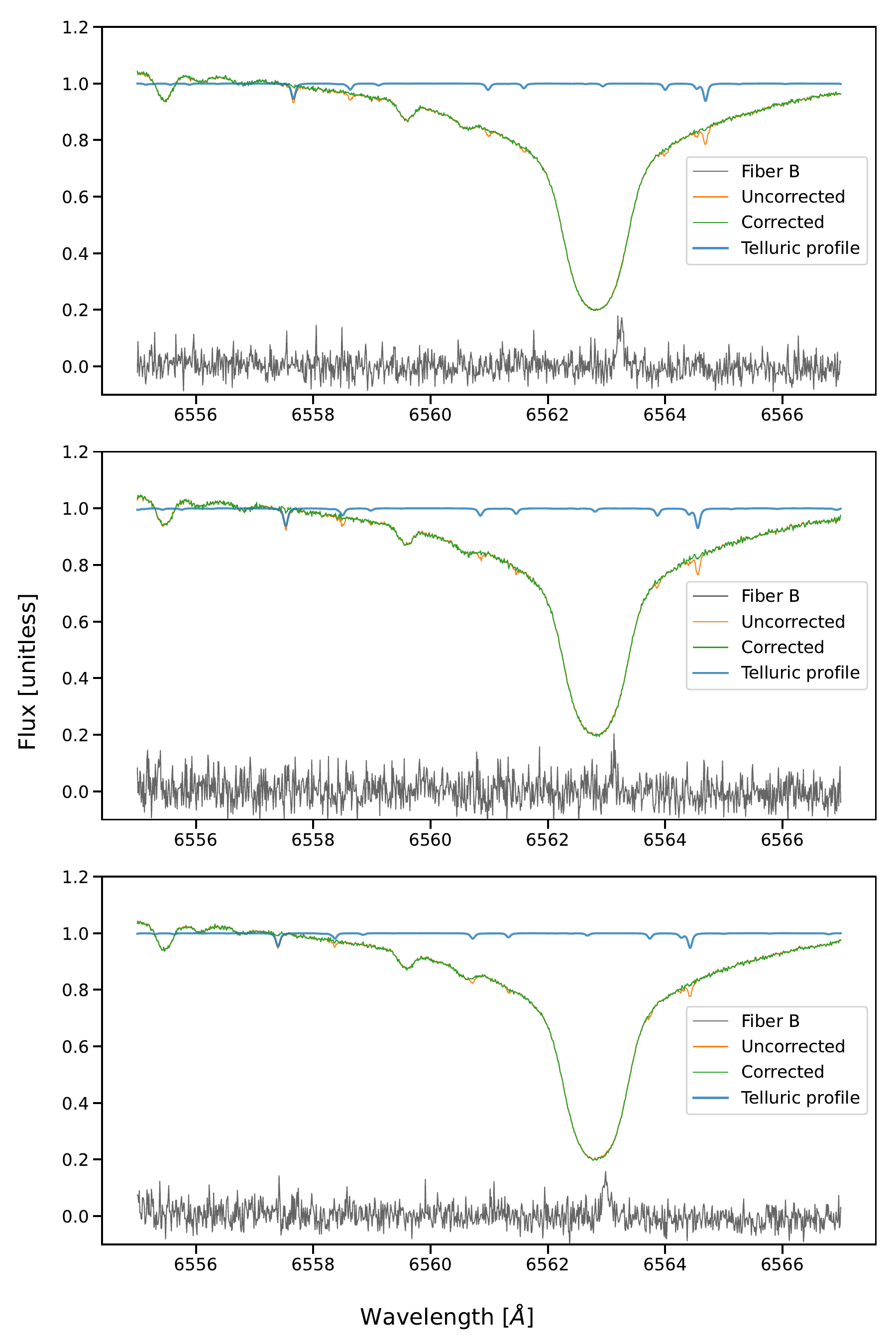}
    \caption{Same as \autoref{fig:earth_correction}, but in region of \Ha. No strong telluric features were detected in this region.}
    \label{fig:earth_correction_ha}
\end{figure}

The first step, as is common in this type of study \citep[e.g.,][]{allart2017, seidel2020b, mounzer2022}, is telluric line correction. We used \texttt{molecfit} \citep{smette2015, kausch2015}, which allows correction for absorption lines due to the Earth's atmosphere in each exposure. To do so, \texttt{molecfit} considers the weather observations and runs a line-by-line radiative-transfer code to calculate a telluric transmission spectrum, such that it reproduces the observed spectra. We refer the reader to the above-mentioned references for more details about the procedure.

As an output, \texttt{molecfit} provides the user with the telluric-corrected spectrum and telluric profile used to correct each spectrum. As the telluric profile does not have an uncertainty, at wavelengths where the atmosphere is almost opaque, the corrected spectra produce features that are not well represented by their uncertainties. To remove them, we identified and replaced pixels with NaN values where the telluric profile was below 50\%, indicating that less than half of the light is successfully transmitted through the Earth's atmosphere. This procedure is relevant for the CCF study and lines falling in the telluric-contaminated regions (e.g., potassium). The effects of telluric correction by \texttt{molecfit} and sky-subtraction done by the DRS are shown in \autoref{fig:earth_correction}. 

After telluric correction, we interpolated the dataset on a common wavelength grid using a cubic spline interpolation. The data were then normalized with a rolling quantile of 85\% and a window of 7 500 pixels. We used a higher quantile than the standard median to more closely represent a stellar spectrum — the spectrum's absorption features (stellar lines) are far more common than emission (mostly cosmic rays), meaning a median will represent the continuum less accurately. We note, however, that the choice of a normalization function is generally irrelevant as long as the method is not variable in time (since we eventually divided spectra by the master-out, the normalization function negates itself). Likewise, we also performed the analysis with a rolling median window to ensure this does not affect the results. The window of 7 500 pixels was chosen empirically, such that the spectrum is mostly flat aside from the noisy blue edge of the detector.

The next step was to shift the telluric-corrected spectra from the solar barycenter rest frame to that of the host star, which was done by considering the systemic velocity and stellar velocity induced by the planet (note that the \texttt{S1D\_SKYSUB} format of the used \texttt{ESPRESSO DRS} is in the solar barycenter, not the Earth's rest frame). There, we calculated the master-out (average out of transit spectrum), again for each night separately, using an unweighted average. We divided the dataset by their respective master-out spectra to remove the stellar contribution to the spectra. 

The spectra are inherently contaminated by the wiggles coming from the Coudé train \citep{allart2020,tabernero2021} optical component in the optical path. These wiggles are currently difficult to correct, and multiple methods have been used to remove them (e.g., local sinusoidal fit \citep{seidel2022}, cubic-spline fit \citep{damasceno2024}, correction on the CCD image before spectrum extraction \citep{sedaghati2021}, or analytical models \citep{bourrier2024}). In this work, we used a similar approach to \cite{damasceno2024}. We took a binned spectrum (a factor of 200 pixels), which we fit with cubic splines. We then divided the spectra by the cubic-spline fit to remove the wiggle contamination.

We then shifted the spectra to the planet's rest frame by assuming the stellar (with a negative sign) and planetary velocity. There, we first focused our analysis on the combined transmission spectrum (average in-transit spectrum), using an unweighted average. Immediately, a strong POLD contamination is visible, as shown in \autoref{fig:transmission_spectrum}. This feature is dominant for most species that are also present in the star. The only exceptions to the lines we probed are lithium (Li), which does not have strong stellar absorption, and the first line of the hydrogen Balmer series (\Ha), which is extremely broad in the stellar spectrum, muting and broadening the effect on the final transmission spectrum. 

We did not detect any species in the transmission spectrum. Instead, we estimated detection limits for the probed lines. We first adjusted the transmission spectrum by $R= 1-\frac{\text{F}_\text{in}}{\text{M}_\text{out}}$, where $\frac{\text{F}_\text{in}}{\text{M}_\text{out}}$ is the in-transit spectrum divided by the master-out.
Then, we created a Gaussian line model for each of the probed lines, with its mean value set to the laboratory line position. The FWHM of the line is set to the instrumental response of ESPRESSO. Finally, amplitude is kept as the sole jump parameter A $\sim$ \( \mathcal{U} \)$(-1,1)$. For POLD-corrected data (\Na\,\Ha\, and K), we ran a MCMC analysis on the transmission spectrum in the 6$\AA$ region centered on the line position. For POLD-uncorrected data, we moved the model mean value to 20 km/s or 1$\AA$. This was done in order to avoid fitting the POLD feature visible in the data, while maintaining the noise behavior in the continuum. The 1$\sigma$ uncertainty was then extracted based on the 68.3\% HDI.

\section{Transmission spectroscopy}
\label{sec:transmission_spectroscopy}
After extracting the transmission spectrum, we checked for resolved line detections, following the list of lines detected by \cite{borsa2021} in WASP-121 b. This list includes lines of Na, H, K, Li, Ca, Mn, and Mg and consists of the easiest lines (detectability-wise) to resolve in the spectra. However, most of these species are likely condensed in the atmosphere of WASP-31 b, owing to its lower temperature compared to WASP-121 b. We focused our endeavor on \Na\ and \Ha. Furthermore, we cross-correlated a forward model generated by \texttt{petitRADTRANS} \citep{molliere2019} to look for Fe. In \autoref{subsec:resolvedlines}, we discuss the transmission spectroscopy of the \Na\ and\Ha\ lines that can be resolved individually. The cross-correlation of iron is discussed in \autoref{subsec:ccf}, and the cross-correlation of CrH is discussed in \autoref{subsec:CrH}.

\subsection{Resolved line-transmission spectroscopy of sodium and hydrogen}
\label{subsec:resolvedlines}
The transmission spectra for \Na\ and \Ha\ are shown in \autoref{fig:transmission_spectrum}. The \Na\ doublet, as most other probed lines, is thoroughly contaminated by POLDs. For these lines, we typically see a negative core (as is the case for HD 209458 b \citep{casasayas-barris2020,casasayas-barris2021}), with weaker positive wings. This makes the detection of the species difficult, as we require forward modeling for both atmosphere and POLD contamination simultaneously to securely distinguish between the signals. We modeled the POLD deformation with EvE \citep{bourrier2013, bourrier2015, bourrier2016}, as discussed in \autoref{sec:RM_modeling_EVE}.

\Ha\ is much broader (and weaker) in the stellar spectrum, causing the POLD distortion to be more spread out and weaker. Furthermore, once combined, the POLDs in the spectral time series cancel each other out, decreasing the amplitude of the deformation. Nevertheless, a visible negative plateau is present, ranging over almost 1 \AA\,(\autoref{fig:transmission_spectrum}). Within it, we see a weak and narrow signal, which is blueshifted roughly by 10kms$^{-1}$. We discuss this further in \autoref{sec:RM_modeling_EVE}.

\subsection{Joint stellar and planet modeling with EvE}
\label{sec:RM_modeling_EVE}
To account correctly for POLDs, we followed the methodology by \cite{dethier2023} using the EvE code. For the simulations, we used values of the system as provided in \autoref{tab:system_parameters}, with $\lambda$ and $v\sin(i_{\star})$ as measured by our analysis (\autoref{tab:RMR_results}). More details about EvE can be found in \cite{bourrier2013, bourrier2016, dethier2023} and \cite{steiner2023}.

\begin{figure}
    \centering
    \includegraphics[width=\columnwidth]{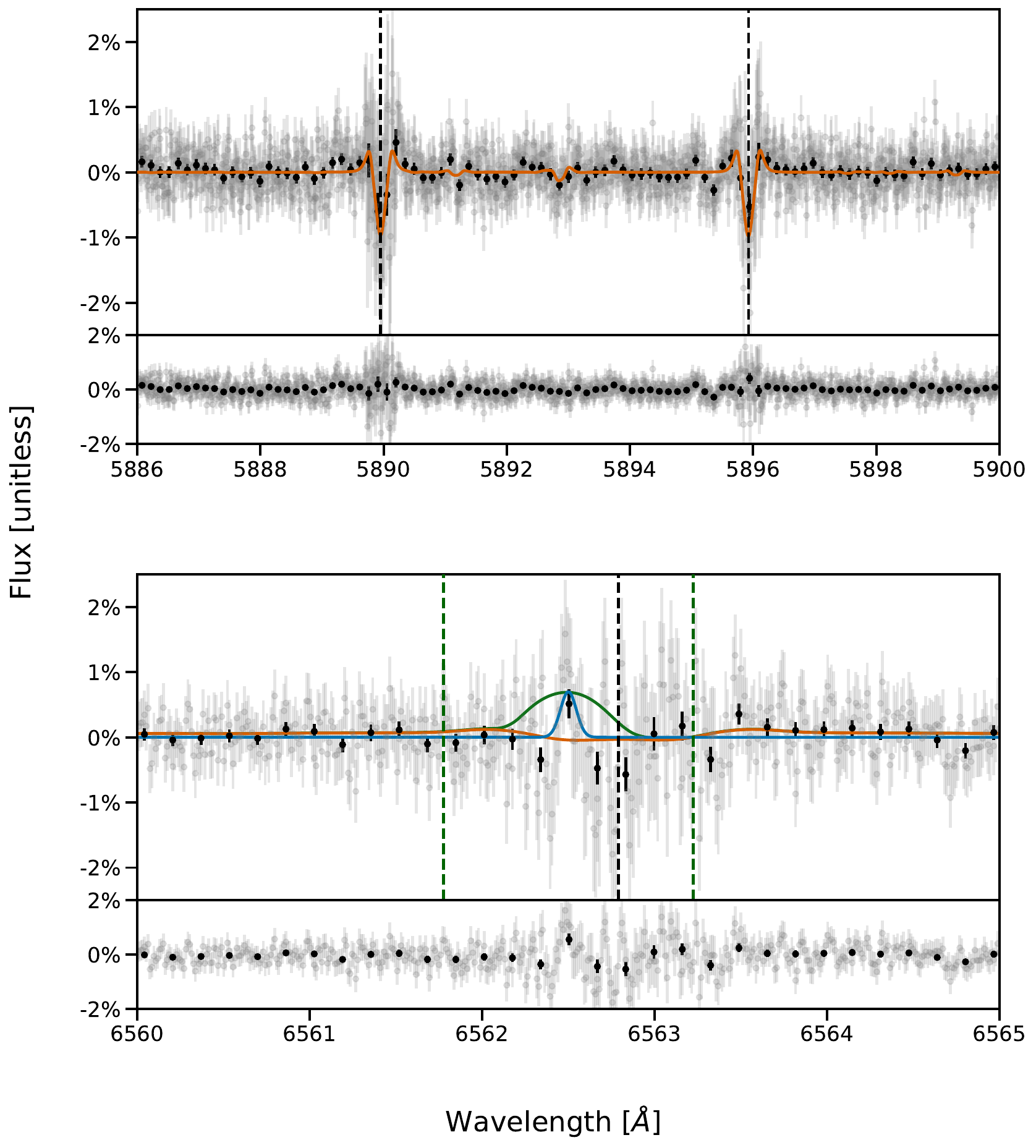}
    \caption{Combined transmission spectrum and its residuals of WASP-31b in the region of sodium doublet (top panels), and Balmer alpha line (bottom panels) in the rest frame of the planet. We plot the full spectrum in gray; that binned by 15x is shown in black. In this notation, absorption goes above zero, and “emission” is below. The laboratory position of the respective lines is shown via dashed black lines. The no-atmosphere model from EvE is plotted as an orange line, and the compound model with the atmosphere is shown as a green line. The vertical dashed green lines correspond to the escape velocity of $\approx$33 km/s, centered on the feature at 6562.5$\AA$. The instrumental response function of ESPRESSO is centered on this feature in blue. Residuals were calculated assuming the no-atmosphere model.}
    \label{fig:transmission_spectrum}
\end{figure}

As no visible signal is detected for sodium, we first simulated a planet disk devoid of an atmosphere. This led to a model that matches the observed signal very well (orange line in top panel; \autoref{fig:transmission_spectrum}). Thus, within our uncertainties, we can fully explain the signal as POLD deformations in the transmission spectra. We show the resulting transmission spectrum and the models in \autoref{fig:transmission_spectrum}.

For the \Ha\,line, due to the hint of a weak signal (possibly due to atmosphere) (\autoref{fig:transmission_spectrum}), we opted for a two-step approach. First, we modeled the transit by simulating a planet disk without any atmosphere. This allows us to show the strength of the POLD effect, which, in this case, is much weaker than in the \Na\ case (orange line in bottom panel; \autoref{fig:transmission_spectrum}). The narrow blueshifted signature seen in the spectra is not explained by this simulation. The width of this feature is close to the instrumental response function (blue line in bottom panel; \autoref{fig:transmission_spectrum}), suggesting it might be an unresolved line. In the second step, we include a realistic atmosphere around the planet to simulate the transit with its atmosphere passing (green line in bottom panel; \autoref{fig:transmission_spectrum}). The simulation leads to a solution in which the atmospheric signature is much broader (albeit of similar strength) than the observed signal as shown in \autoref{fig:transmission_spectrum}. 

Neither of these models fully explains the observation. In particular, neither can clearly explain the amplitude of the negative core of the POLD distortion. The “no-atmosphere” model (orange line in bottom panel; \autoref{fig:transmission_spectrum}) for \Ha\ underestimates the amplitude by a factor of $\sim$10. This model does not reproduce the feature observed at 6562.5 $\AA$, for which we attempted to use an atmosphere model in EvE. The modeled atmospheric signature (green line in bottom panel; \autoref{fig:transmission_spectrum}) is much wider than what we observed, and, as such, we attempted to match the width of the line by decreasing the input temperature in EvE. Lowering the equilibrium temperature to 750 K (half of the value reported by \cite{anderson2011}) did not significantly affect the width of the modeled atmospheric line. The atmospheric density profile is computed with a hydrostatic profile. The temperature profile is isothermal. Day-to-night-side wind was applied as a bulk wind to the atmosphere. Rotation of the planet and the atmosphere were indeed taken into account, considering the planet tidally locked. Therefore, the included broadening of the line is due to the temperature and rotation of the atmosphere. We observe this feature on all nights (\autoref{subsec:halpha-feature}). Given the instrumental response function, the observed feature, if real, might be an unresolved line, potentially an alias line with the Balmer alpha. Within the region of 6562–6563 $\AA$, the potential candidates, except for \Ha, are V II (6562.4 $\AA$), Li II (6562.61 $\AA$), Th I (6562.647 $\AA$), and Au I (6562.68 $\AA$).

\begin{table}[]
    \centering
    \caption{Upper limits for \Na\ and \Ha with the corresponding height in planetary radii in parentheses.}
    \begin{tabular}{lr}
    \toprule
    Line & 3$\sigma$ upper limit \\
    \midrule
    $\text{Na-D2} $ & $ 1.962\%$ $ (1.48\,R_p)$ \\
    $\text{Na-D1} $ & $ 1.311\%$ $ (1.34\,R_p)$ \\
    $\text{H$\alpha$} $ & $ 1.689\%$ $ (1.42\,R_p)$ \\
    \bottomrule
    \end{tabular}

    \label{tab:upper-limits}
\end{table}

As other species (aside from \Na\ and \Ha) were outside the main scope of this paper, we did not simulate the POLD effect on the rest of the species (\autoref{subsec:transmission-other-lines}). For Li, we do not see POLD deformation, as the spectral line does not appear in the stellar spectrum. For other lines, we see strong negative features, similar to the \Na\ POLD deformations.

\subsection{Cross-correlation function of iron}
\label{subsec:ccf}
The iron template model was generated using the system parameters, as reported by \autoref{tab:system_parameters}, assuming the Guillot T-P profile \citep{guillot2023} with pressures between $[10^{-10}; 10^{2}]$ bar. We assumed weighting by the pixel's uncertainty, that is\footnote{Note we do not use the global SNR here. This is to ensure locally noisy pixels do not contaminate the resulting CCF.}: 

$$w_i=\frac{1}{\sigma_{i}^2}$$,

where the CCF calculation follows \cite{hoeijmakers2019} with a few modifications due to weighting and a non-normalized template. 
$$\text{CCF} (t,v) =\frac {\sum_{i}^N x_i(t) T_i(v) w_i(t)}{\sum_{i}^N T_i(v) w_i(t)}$$,

here, the i corresponds to each pixel of the spectrum, with its respective weight, $w_i$. The template $T_i(v)$ is the previously generated model from petitRADTRANS, and $x_i(t)$ is the spectrum at time t. 

\begin{figure}[!h]
    \centering
    \includegraphics[width=\columnwidth]{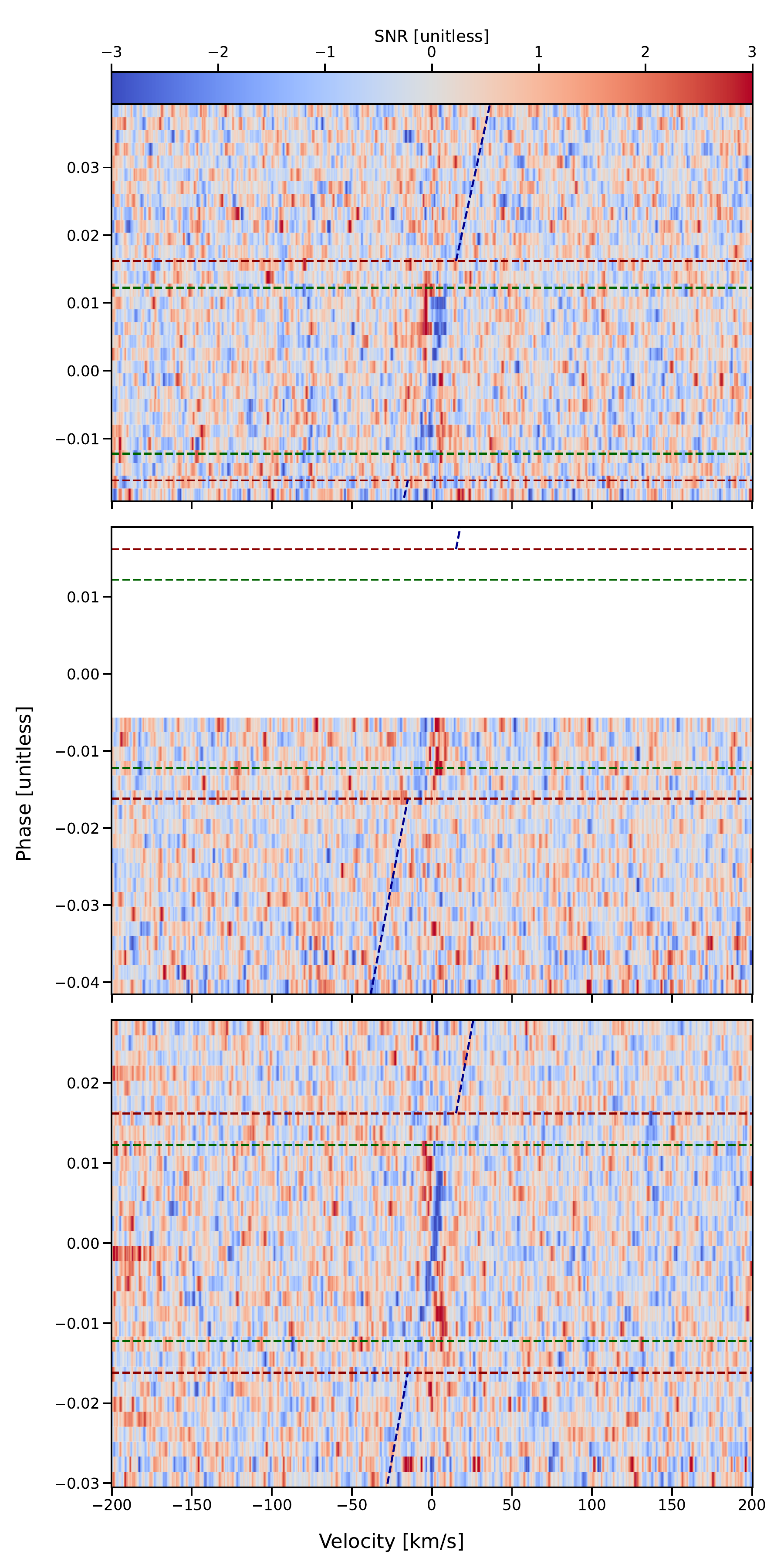}
    \caption{Phase-resolved cross-correlation of iron template with the dataset of WASP-31b, in the rest frame of the star. Each panel corresponds to a separate night (top panel for first night, middle for second, and bottom for third night). The contact points T1-T4 are plotted as red horizontal dashed lines, with contact points T2-T3 shown as horizontal dashed green lines. The planetary track is shown as a dashed line blue for the out-of-transit section. Absorption features are shown in red (positive) and negative features in blue.}
    \label{fig:CCF_iron_phase}
\end{figure}

\begin{figure}[!h]
    \centering
    \includegraphics[width=\columnwidth]{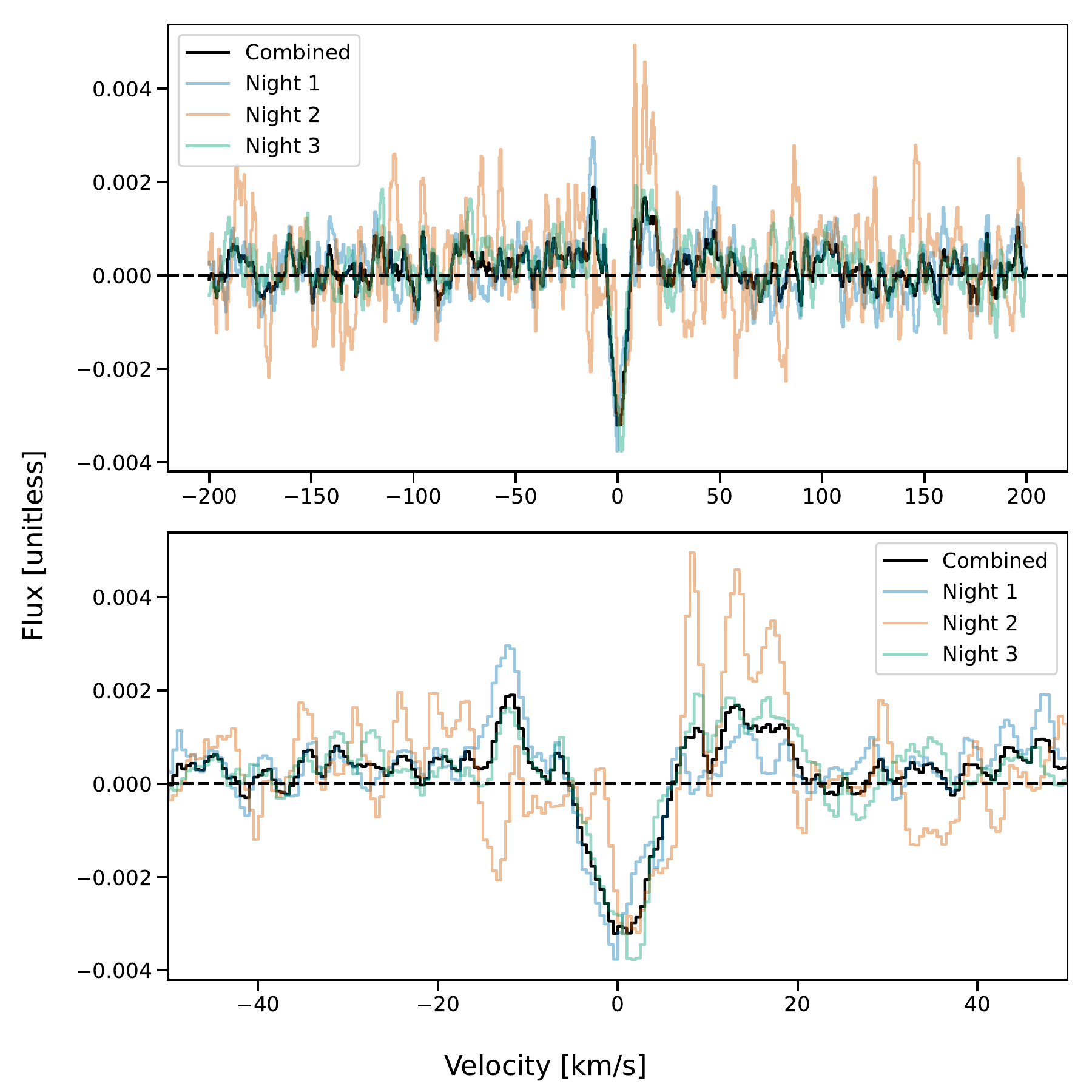}
    \caption{Cross-correlation of iron template with the combined and night-specific transmission spectroscopy in the rest frame of the planet. The combined CCF of all nights is shown in black, with the nights shown separately in blue (night 1), orange (night 2), and green (night 3). The top panel shows the full velocity range over which the CCF was calculated. The bottom panel shows a zoomed-in view of -50,50 kms$^{-1}$. Absorption is positive with respect to the continuum.}
    \label{fig:CCF_iron_combined}
\end{figure}

We show the resulting phase-resolved CCF in \autoref{fig:CCF_iron_phase}, and the combined 1D CCF in \autoref{fig:CCF_iron_combined}. We see a strong contamination from the POLD effect, with no atmospheric signature within. As shown in \autoref{fig:CCF_iron_phase}, the planetary track is almost unresolvable from the Doppler shadow. 

\subsection{Cross-correlation function of chromium hydride}
\label{subsec:CrH}
WASP-31b is the first exoplanet on which a chromium-hydride (CrH) detection has been noted, first tentatively using HST STIS/WFC~\citep{braam2021}, and later with high-resolution spectroscopy using GRACES and UVES~\citep{flagg2023}. 
While CrH only has negligible opacity throughout most of the optical bandpass, it does have some weak spectral lines redward of $\sim$690\ nm and its main molecular bands starting at $\sim$765\ nm. While this leaves only a small portion of the ESPRESSO detector (380 -- 780\,nm) overlapping with CrH lines, in light of the previous detections, we nevertheless opted to search for CrH absorption in the transmission time series of WASP-31b. 

\begin{figure}[!h]
    \centering
    \includegraphics[width=\columnwidth]{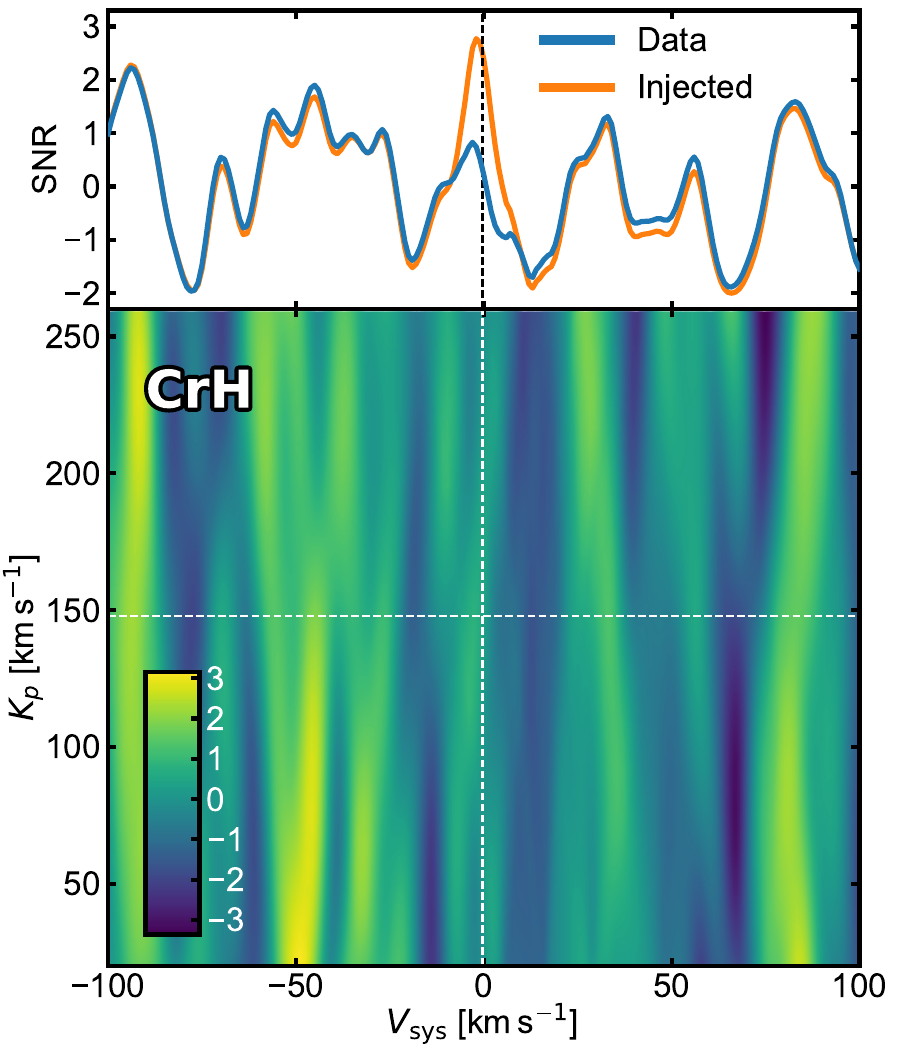}
    \vspace{-4mm}
    \caption{Chromium-hydride (CrH) cross-correlation search and sensitivity check.  The bottom panel shows the phase-folded $K_p$-$V_{\mathrm{sys}}$ signal-to-noise map of the ESPRESSO WASP-31b transit data cross-correlated with a CrH model template.  The dashed lines denote the expected values of ~\citep[$V_{\mathrm{sys}}$ = $-$0.1\,km\,s$^{-1}$, and $K_p$ = 148.0\,km\,s$^{-1}$,][]{gaiadrs3, flagg2023}. The top panel shows a one-dimensional slice at the expected $K_p$ (along the horizontal dashed white line) for both the data alone (blue) and the data with an injected model (orange). The difference between the blue and orange lines is a measure of the data sensitivity to an underlying signal. In this case, these data only have limited sensitivity to CrH, which has its strongest opacity band heads redward of the ESPRESSO wavelength coverage. Although no clear CrH signal is observed in the data here, we cannot conclusively confirm or rule out its presence in the atmosphere of WASP-31b due to limited sensitivity.
    }
    \label{fig:CrH-kpvsys}
\end{figure}

As the high-resolution detection of CrH by \cite{flagg2023} was made from a data detrending method based on singular-value decomposition, we opted to use a similar approach, in contrast to the \texttt{molecfit}-based telluric correction used for our main analysis. For this, we processed the data using the principal-component-analysis (PCA) based detrending framework of \cite{pelletier_where_2021, pelletier_vanadium_2023, pelletier_crires_2024, bazinet_subsolar_2024}. In short, this algorithm processes each spectral order of each transit time series, wherein all spectra are corrected for bad pixels, aligned to a common continuum level, corrected for out-of-transit median spectra, and divided by a five-principal-component reconstruction of the data.
The analysis procedure closely follows \cite{pelletier_crires_2024}, to which we refer the reader for more details. We also cross-correlated the CrH template following our main \texttt{molecfit}-based approach as described in \autoref{subsec:tsextraction} and \autoref{subsec:ccf}, finding compatible results.

For the cross-correlation, we generated a fiducial solar composition~\citep{asplund2009} chemical-equilibrium transmission template of WASP-31b using the \texttt{SCARLET} atmosphere modeling framework~\citep{benneke_atmospheric_2012, benneke_how_2013, benneke_strict_2015, benneke_sub-neptune_2019} (by default, \texttt{petitRADTRANS} does not provide CrH line lists at high spectral resolution).  
The CrH cross-sections used in the model were computed from the line list of \cite{burrows_new_2002} using \texttt{HELIOS-K}~\citep{grimm_helios-k_2015, grimm_helios-k_2021}. Chemical-equilibrium abundance profiles were calculated using \texttt{FastChem}~\citep{stock_fastchem_2018, stock_fastchem_2022} assuming the thermal structure of the atmosphere of WASP-31b to be isothermal at an equilibrium temperature of 1393\,K~\citep{braam2021} and having a gray cloud deck at 100\,mbar. CrH has the advantage that, unlike Na and other neutral metals, it is not expected to be present in the stellar photosphere and hence should not be contaminated by the POLD effect.

We then cross-correlated the CrH transmission template, once with the detrended data, and once with the detrended data after the generated WASP-31b transmission template had been injected in the transit time series (before detrending) on top of the real signal.  This latter scenario of injecting and recovering a realistic model of the atmosphere of WASP-31b in the data serves as a sensitivity test of how strongly our observations should detect an underlying real signal. While we do not find any evidence of CrH absorption in the ESPRESSO transit data, based on the injection-recovery test, we also likely would not have been sensitive to a real signal (\autoref{fig:CrH-kpvsys}).  We therefore cannot confirm (or refute) the CrH detection on WASP-31b, largely owing to the coverage of ESPRESSO being limited to wavelengths blueward of 780\ nm. In particular, we note that the CrH detection made by \cite{flagg2023} used the 860–895\ nm wavelength range not covered by our data. 

\section{Discussion}
\label{sec:discussion}
We confirm the measurement of the sky-projected spin-orbit angle of WASP-31b, as shown by \cite{brown2012}. Our precision suggests almost perfect projected alignment; however, as shown in \cite{attia2023}, the true spin-orbit angle is likely to be underestimated. An analysis of the photometry did not lead to stellar rotation-period measurement needed for the measurement of $i_{\star}$ and the derivation of the 3D spin-orbit.

The non-detection of the resolved lines \Na, K, Mg, Ca, and Mn is expected, as the contamination by the POLD effect hides any potential signal. We modeled the POLDs with a cutting-edge model to assess that the low-significance feature observed at \Ha\ does not originate from POLDs. Li is an exception to the list of lines polluted by POLDs, as the host star does not exhibit significant absorption of Li, which removes any contamination by the POLD effect. 

\begin{figure}[!h]
    \centering
    \includegraphics[width=\linewidth]{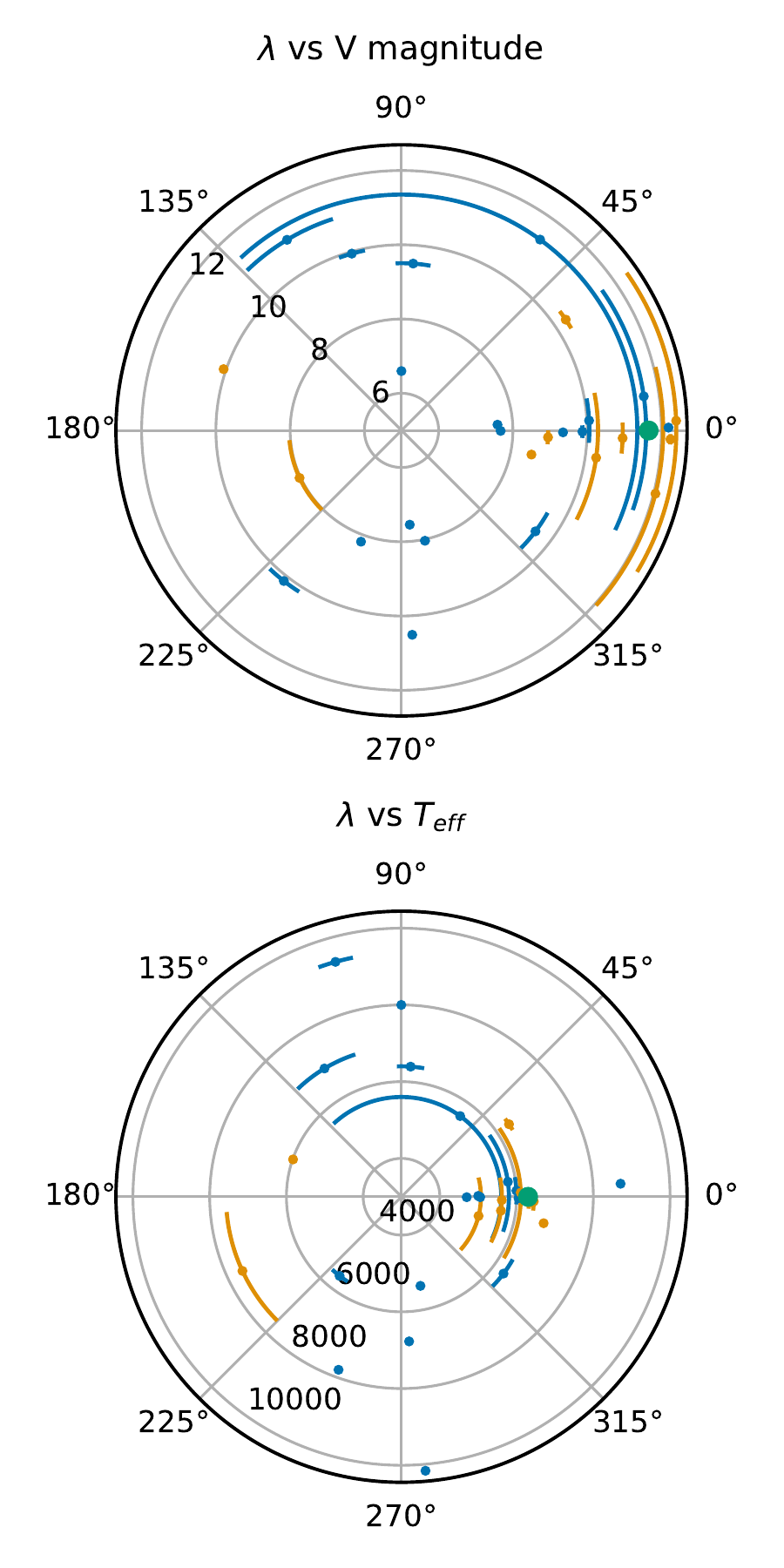}
    \caption{Detections (blue) and non-detections (orange) of \Na\ doublet lines using high-resolution spectrographs based on projected spin-orbit angles ($\lambda$) of the system. The green point is the position of WASP-31 b found using our observed values for $\lambda$. The top panel shows the (non-)detections with respect to V magnitude. The bottom panel shows the (non-)detections with respect to $T_{eff}$ of the host star. Data were extracted on 7 March, 2025. Although sodium detections exist spanning vastly different orbital alignments, non-detections show a clear preference for aligned systems (1st and 3rd quadrants); however, given the small number of non-detection publications and the higher chance of close-in giants being aligned, we cannot distinguish them from observational bias.}
    \label{fig:sodium_detections}
\end{figure}

We illustrate the difficulty of achieving atmospheric detections due to POLD effects in \autoref{fig:sodium_detections}, showing the (non-)detections of sodium, depending on alignment. As shown, misalignment almost ensures an atmosphere's detectability, as the POLD contamination can be distinguished from the planetary track. The majority of the planets with sodium detection claims are for hot and ultra-hot Jupiters, for which a significant amount of data exists, including high-resolution spectrographs. We note, however, that there is likely observational bias due to a low number of non-detection publications. In WASP-31 b, the combination of system parameters leads to the orbital track being undistinguishable from the Doppler shadow, as illustrated in \autoref{fig:CCF_iron_phase}.

The case of \Ha\ is quite unique, as it is the only species present in the stellar spectrum not visibly contaminated by the POLD effect. This is likely due to the fact that the POLDs effect scales with the stellar line gradient and is muted for very broad lines such as \Ha. Indeed, modeling of the POLDs effects shows a small amplitude deformation. Inside the transmission spectrum, we observe a low S/N feature of unknown origin, either due to possible detection of \Ha, or due to noise in the spectra. As the significance of the feature is below 3$\sigma$, we do not claim any detection from the present data.

In the CCF regime, we looked for iron as the main species within the ESPRESSO wavelength range. However, Fe is severely contaminated by POLDs, disallowing any form of detection limit extraction. On the other hand, the non-detection of iron is indicative of a cloud deck, which is expected for these kinds of systems. 

Finally, we attempted to search for CrH in our data, which had been suggested by \cite{braam2021} and then confirmed by \cite{flagg2023}. For methodology reasons, we used a different pipeline in order to align our work with the methodology followed by \cite{flagg2023}. However, as can be seen from \autoref{fig:CrH-kpvsys}, we are unable to detect expected signal from the injection-recovery test, which is likely due to the low sensitivity of CrH in the wavelength range of ESPRESSO. Thus, we make no claim on the presence of CrH in the atmosphere of WASP-31 b.

These results highlight the difficulty of making high-resolution observations for aligned exoplanets, which could suggest that a better observing strategy for high-resolution spectroscopy might be needed, particularly in light of the future ELT/ANDES observations of terrestrial planets. This could motivate a two-step approach of using high-resolution spectrographs at smaller telescopes to observe the projected spin-orbit angle, applying a filter to the target selection based on the overlap of the planetary track with the core POLD contamination. Alternatively, the use of models simulating both the planetary disk and the atmosphere, as EvE does, could be used to detect and characterize signatures when the planetary tracks are unresolved from the Doppler shadow.

\section{Conclusions}
\label{sec:conclusion}
In this work, we analyzed the ESPRESSO dataset of WASP-31b, an inflated hot Jupiter. We first analyzed the RM signature using the revolutions method \citep{bourrier2021}, which allowed us to constrain the projected obliquity to \obliquityresult\ and projected stellar rotation to \vsiniresult, increasing the precision by factors of ten and seven, respectively, with regard to previous works \citep{brown2012}.

Our analysis shows significant contamination of the transmission spectra by the POLD effect, which effectively disallows the analysis of lines that are present in the star. The only exceptions are the broadened \Ha, where the POLD signature is much weaker due to the stellar line broadness, and Li, which does not show a stellar line. The POLD contamination is also visible in the CCF regime, where we looked for iron. 

Finally, as WASP-31b is the first planet with CrH detection \citep{braam2021,flagg2023}, we tried to perform a CCF with this species as well. We do not detect any CrH; however, using an injection-recovery test, we show that we are unable to extract reliable detection with this dataset, in particular due to the wavelength range of ESPRESSO not overlapping with strong CrH lines.

This paper highlights the need to distinguish the planetary track from the core of the POLDs effect, which is one of the core issues in transmission spectroscopy at the advent of the ELT observations.

\begin{acknowledgements}
This project has received funding from the European Research Council (ERC) under the European Union's Horizon 2020 research and innovation programme (project {\sc Four Aces}; grant agreement No 724427). MS and VB's work has been carried out in the frame of the National Centre for Competence in Research PlanetS supported by the Swiss National Science Foundation (SNSF). MS and VB acknowledge funding from the European Research Council (ERC) under the European Union's Horizon 2020 research and innovation programme (project {\sc Spice Dune}; grant agreement No 947634). It has also been carried out in the frame of the National Centre for Competence in Research PlanetS supported by the Swiss National Science Foundation (SNSF). MS, DE and VV acknowledges financial support from the SNSF for project 200021\_200726. Co-funded by the European Union (ERC, FIERCE, 101052347). Views and opinions expressed are however those of the author(s) only and do not necessarily reflect those of the European Union or the European Research Council. Neither the European Union nor the granting authority can be held responsible for them. This work was supported by FCT - Funda\c{c}\~{a}o para a Ci\^{e}ncia e a Tecnologia through national funds by these grants: UIDB/04434/2020 DOI: 10.54499/UIDB/04434/2020, UIDP/04434/2020 DOI: 10.54499/UIDP/04434/2020. R.A. acknowledges the Swiss National Science Foundation (SNSF) support under the Post-Doc Mobility grant P500PT\_222212 and the support of the Institut Trottier de Recherche sur les Exoplan\`{e}tes (iREx). This work has been carried out within the framework of the National Centre of Competence in Research PlanetS supported by the Swiss National Science Foundation. The authors acknowledge the financial support of the SNSF. The INAF authors acknowledge financial support of the Italian Ministry of Education, University, and Research with PRIN 201278X4FL and the ``Progetti Premiali'' funding scheme. JIGH, ASM, CAP, and RR acknowledge financial support from the Spanish Ministry of Science, Innovation and Universities (MICIU) projects PID2020-117493GB-I00 and PID2023-149982NB-I00. This work was financed by Portuguese funds through FCT (Funda\c{c}\~{a}o para a Ci\^{e}ncia e a Tecnologia) in the framework of the project 2022.04048.PTDC (Phi in the Sky, DOI 10.54499/2022.04048.PTDC). CJM also acknowledges FCT and POCH/FSE (EC) support through Investigador FCT Contract 2021.01214.CEECIND/CP1658/CT0001 (DOI 10.54499/2021.01214.CEECIND/CP1658/CT0001). We acknowledge financial support from the Agencia Estatal de Investigaci\'{o}n of the Ministerio de Ciencia e Innovaci\'{o}n MCIN/AEI/10.13039/501100011033 and the ERDF ``A way of making Europe'' through project PID2021-125627OB-C32, and from the Centre of Excellence ``Severo Ochoa'' award to the Instituto de Astrof\'{\i}sica de Canarias. A.P. acknowledges grants from the Spanish program Unidad de Excelencia Mar\'{\i}a de Maeztu CEX2020-001058-M, 2021-SGR-1526 (Generalitat de Catalunya), and support from the Generalitat de Catalunya/CERCA. This work has been carried out within the framework of the NCCR PlanetS supported by the Swiss National Science Foundation under grants 51NF40\_182901 and 51NF40\_205606. FPE and CLO would like to acknowledge the Swiss National Science Foundation (SNSF) for supporting research with ESPRESSO through the SNSF grants nr. 140649, 152721, 166227, 184618 and 215190. The ESPRESSO Instrument Project was partially funded through SNSF's FLARE Programme for large infrastructures.

\\
Apart from the already mentioned software, this project used Python (version 3.10.12) and its respective implemented libraries. Furthermore, these external packages were also utilized:
astropy \citep{astropycollaboration2022}, specutils, numpy, scipy, matplotlib and pandas, mpire, pyvo, seaborn and bokeh.

\end{acknowledgements}
\clearpage

\bibliographystyle{aa} 

\bibliography{references} 
\newpage
\clearpage

\begin{appendix}
\section{Observing conditions for spectroscopic exposures}
\begin{figure}[h]
\centering
\includegraphics[width=.8\columnwidth]{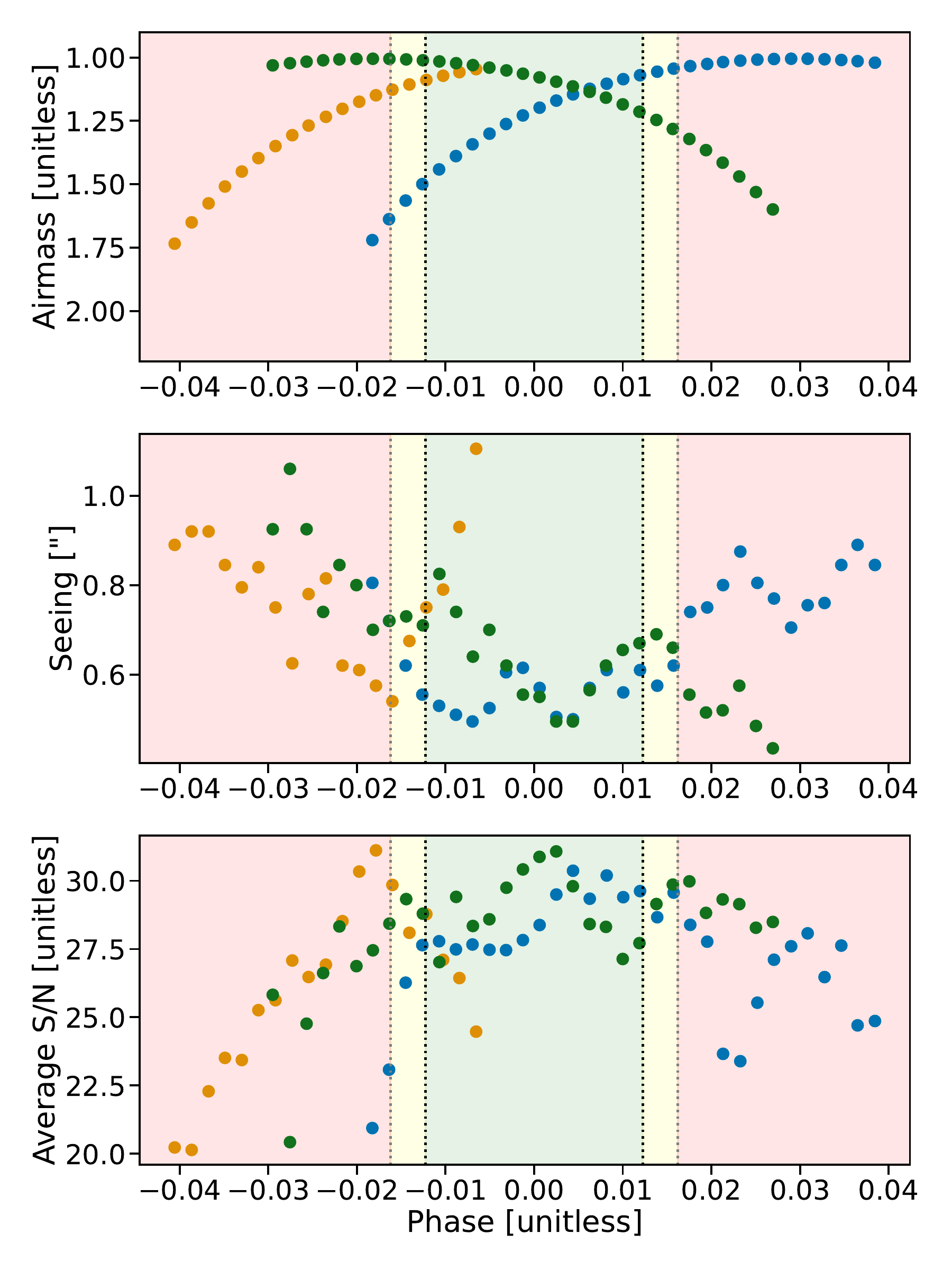}
  \caption{Observation log of the three nights ESPRESSO observation. The red, yellow and green regions correspond to out-of-transit, partial-transit and full-transit phases, with the transit contact points plotted as gray ($T_1$,$T_4$) or black ($T_2$, $T_3$) dotted lines. The blue, orange and green points corresponds to first, second and third night, respectively. Top panel is showing the Airmass variation for each night in the phase space. Middle panel is the same plot but for seeing variation, and bottom panel is the average S/N during the observations. The quality of nights is comparable for the ESPRESSO nights. During the second night (orange points) of ESPRESSO, the seeing condition got worse mid-transit, which led to stopping the observation.}
     \label{fig:observation_log}
\end{figure}

In \autoref{fig:observation_log} we show the evolution of the observing conditions for each night. 

\section{Ephemeris choice}
\label{subsec:ephemeric-choice}
We would like to note that during our analysis of RM, we also used the ephemeris by \cite{patel2022}. However, this set is clearly incompatible with the observations and biases the result. The transit window centers calculated from \cite{kokori2023} ($1\sigma \simeq 0.5 min$) and \cite{patel2022} ($1\sigma \simeq 2.5 min$) differ by 10 minutes at the night \#1 transit, as obtained by NASA Exoplanet Archive Transit Service. 

This discrepancy arrives from the difference in orbital period ($\simeq$1.9 s). The orbital period by \cite{kokori2023} ($3.405\,887\,50\pm0.000\,000\,27$ d) and the orbital period by \cite{patel2022} ($3.405\,909\,5^{+0.000\,004\,7}_{-0.000\,004\,8}$ d) are incompatible by around $4.5\sigma$. Given the $T_c$ of \cite{kokori2023} is 375 transit events earlier than $T_c$ of \cite{patel2022}, this already propagates to $\simeq 712$ s difference. 
	
\begin{figure}
    \centering
    \includegraphics[width=\linewidth]{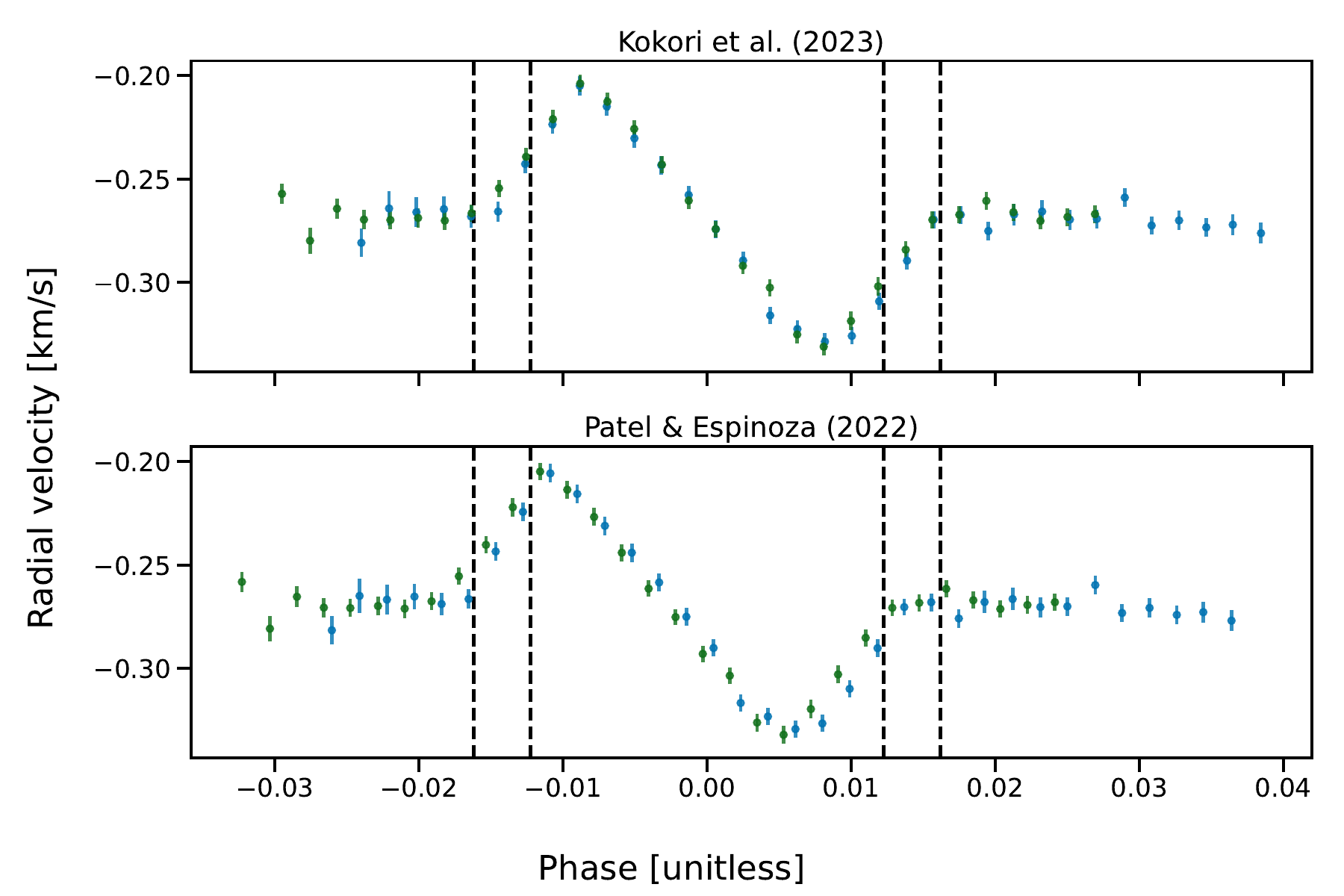}
    \caption{RV curve based on ephemeris provided by \cite{kokori2023} (top) and \cite{patel2022} (bottom). The two nights are visually horizontally well aligned in the top panel, while there is clear misalignment in the bottom panel. For visual purposes, only night 1 (blue) and night 3 (green) are plotted. Contact points are plotted as dashed vertical lines in black. RV are corrected for the stellar velocity induced by the planet.}
    \label{fig:rv-ephemeris-comparison}
\end{figure}

Visual comparison between the RV curve assuming \cite{kokori2023} and \cite{patel2022} shows a clear offset between the first and last night of our data assuming ephemeris by \cite{patel2022}, which is shown in \autoref{fig:rv-ephemeris-comparison}. Our ephemeris refinement is also compatible with \cite{kokori2023}, and is more precise at the transit centers for all observed nights, and we opt to use our refined ephemeris in the analysis.

\clearpage

\section{Rossiter-McLaughlin analysis}
\subsection{Posterior of individual intrinsic CCF}
\label{subsec:pdf_intrinsic}
\begin{figure*}[h]
    \centering
    \includegraphics[width=1.9\columnwidth]{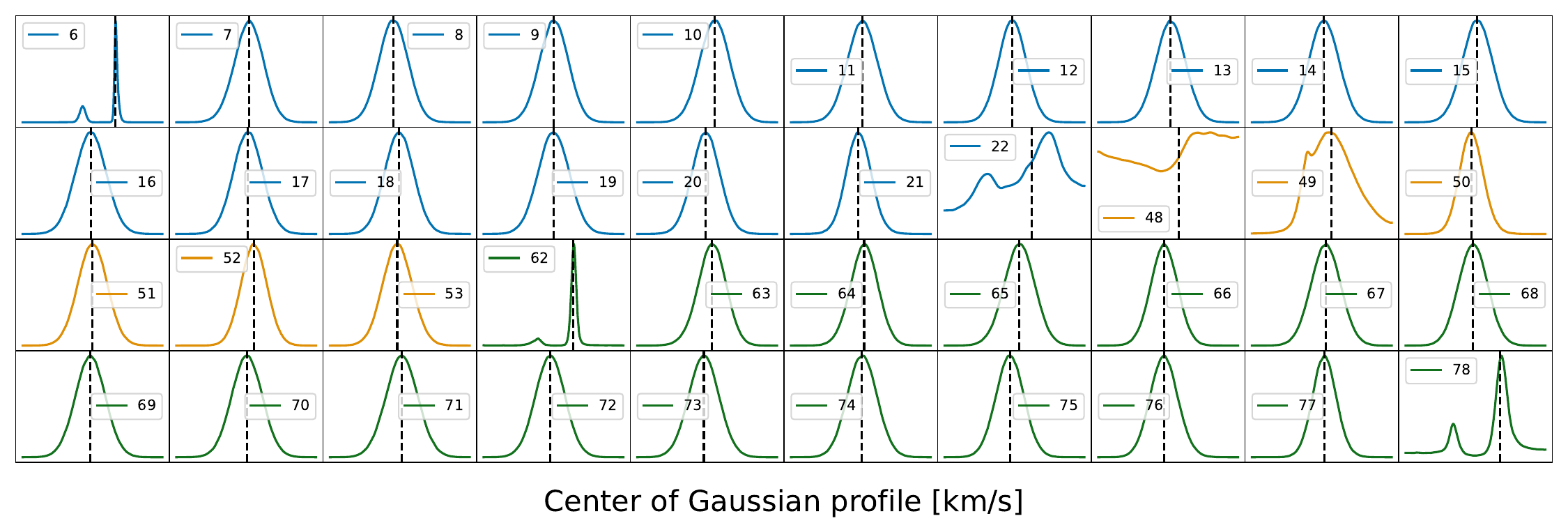}
    \includegraphics[width=1.9\columnwidth]{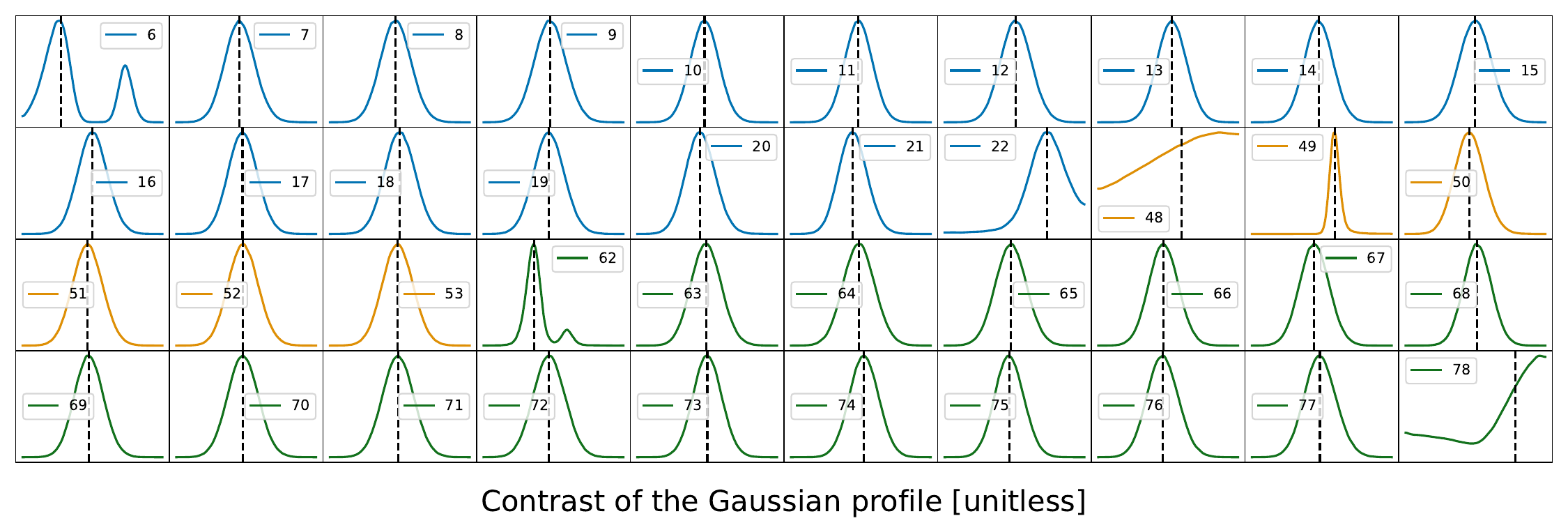}
    \includegraphics[width=1.9\columnwidth]{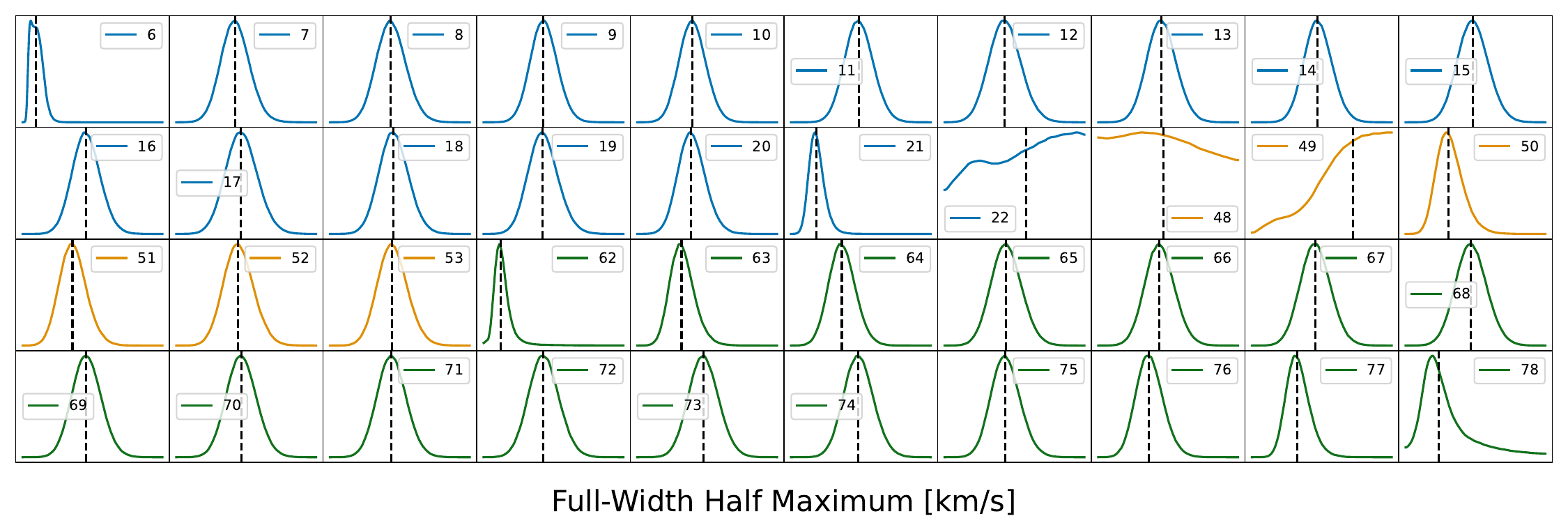}
    \caption{Posteriors distribution of the analysis of intrinsic CCF's. The top panel shows the center of the Gaussian, the middle panel is the amplitude (contrast) of the Gaussian and the bottom panel is the FWHM. The distributions are color-coded by night (blue for \#1, orange for \#2 and green for \#3). The label corresponds to a specific index assigned to each spectrum at the start of the pipeline, after the spectrum list has been sorted by time (meaning these are phases sorted per night). Only in-transit data are shown (between T1 and T4 contact points). As we use this plot only for illustration of how well/poorly defined the PDFs are, the x and y-axis ticks were removed. The distributions are well-behaved for all spectra during the full transit, but as expected, the first and last transiting spectrum show poorly defined distributions, and were thus removed from the analysis. Furthermore, the second spectrum in second night (\#49) shows poor FWHM distribution, and has also been removed from the analysis.}
    \label{fig:posteriors_intrinsicccf}
\end{figure*}

The posterior distributions of fitting individual intrinsic CCF profiles with a Gaussian model. As shown in \autoref{fig:posteriors_intrinsicccf}, there are several exposures which are poorly defined. Note that only in-transit spectra are calculated, so out-of-transit indices are missing. These were removed from the final analysis. The full list of removed spectra is: \#6; \#22; \#48; \#49; \#62 and \#78. Filtering these have marginal impact on the analysis, with differences much smaller than 1$\sigma$ of the resulting values.

\clearpage
\newpage
\subsection{Contrast anomaly in raw CCF}
\label{subsec:contrast-anomaly}
\begin{figure}[h]
    \centering
    \includegraphics[width=\columnwidth]{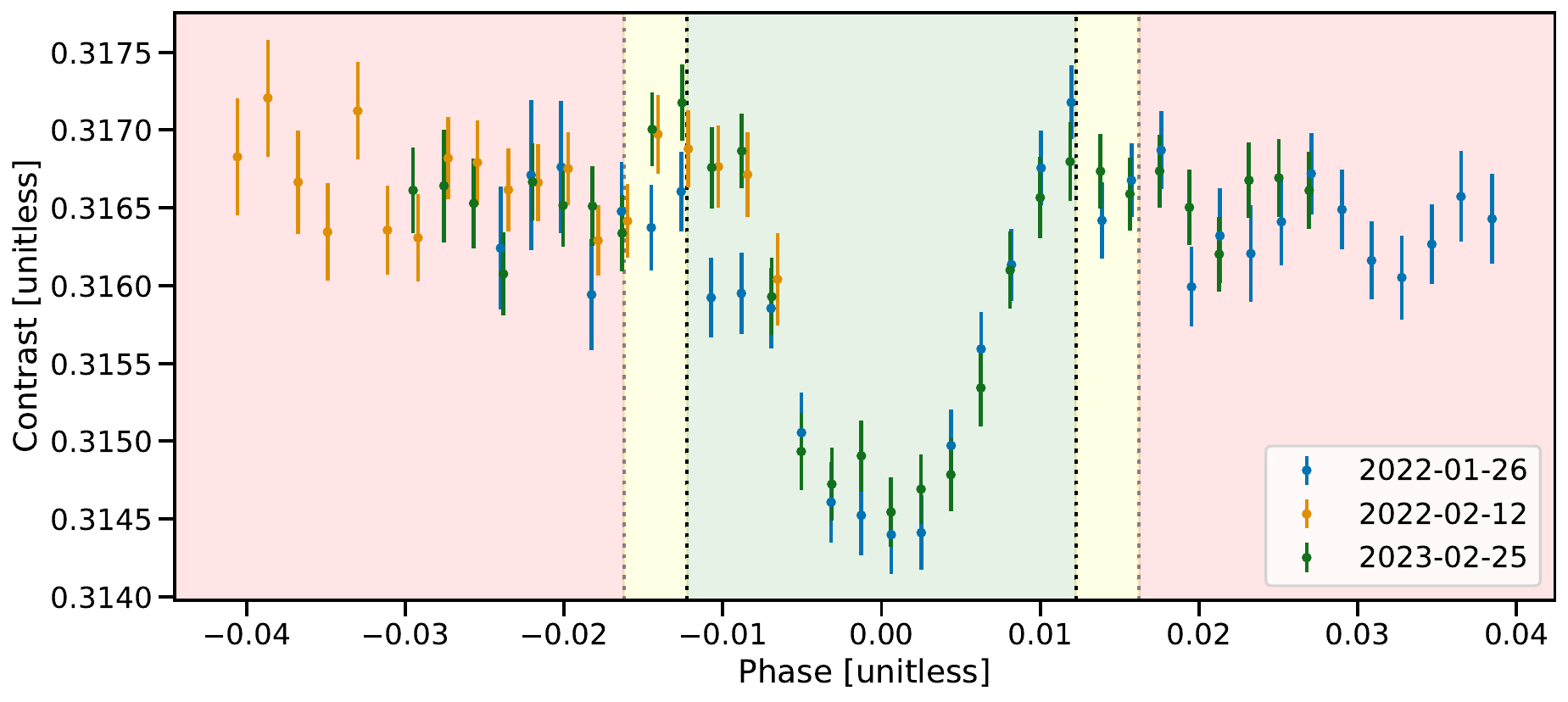}
    \caption{Contrast of the raw CCF, extracted by fitting a Gaussian profile. Each night is color-coded (blue, orange, and green for first, second and third night, respectively). The first few in-transit exposures of the first night are clearly deviating from the trend. The red, yellow and green regions correspond to out-of-transit, partial-transit and full-transit times, with the transit contact points plotted as gray ($T_1$,$T_4$) or black ($T_2$, $T_3$) dotted lines.}
    \label{fig:contrast-per-night}
\end{figure}

As shown in \autoref{fig:contrast-per-night}, we see a deviation in the contrasts from the expected shape for the first night, first 4 in-transit exposures. While we expect a trend to appear due to the transit itself, a few of the first exposures in the first night, namely \#6-\#9, deviate from the expected trend, likely due to faculae occultation. We remove these exposures from the analysis, as these are clearly not well characterized with our model.

\twocolumn

\section{Other strong lines in the transmission spectra}
\label{subsec:transmission-other-lines}
\begin{figure}
    \centering
    \includegraphics[width=\linewidth]{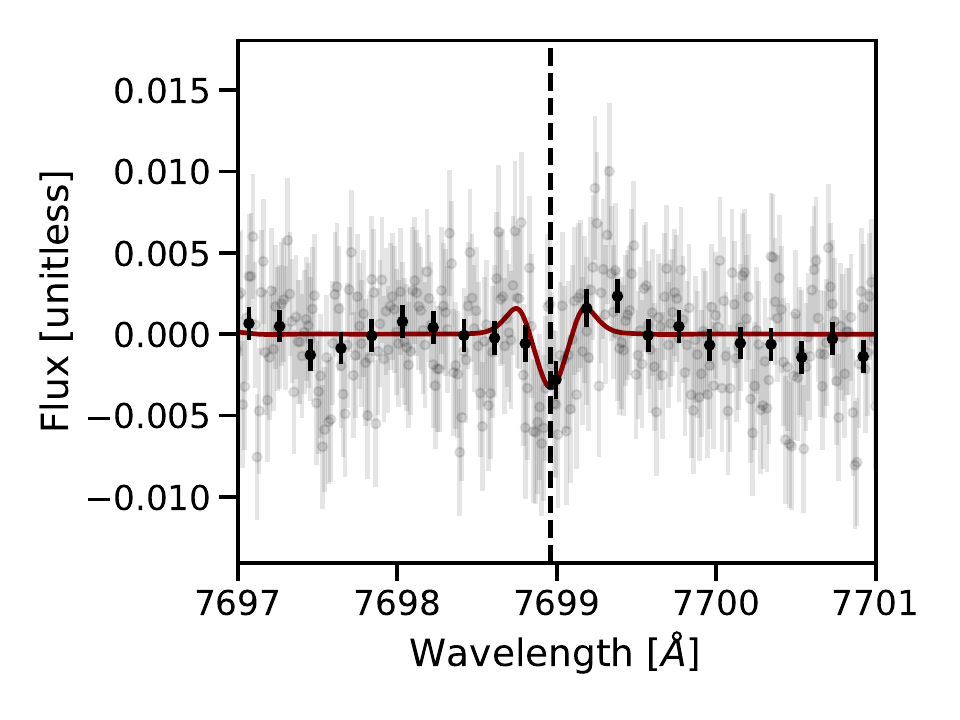}
    \includegraphics[width=\linewidth]{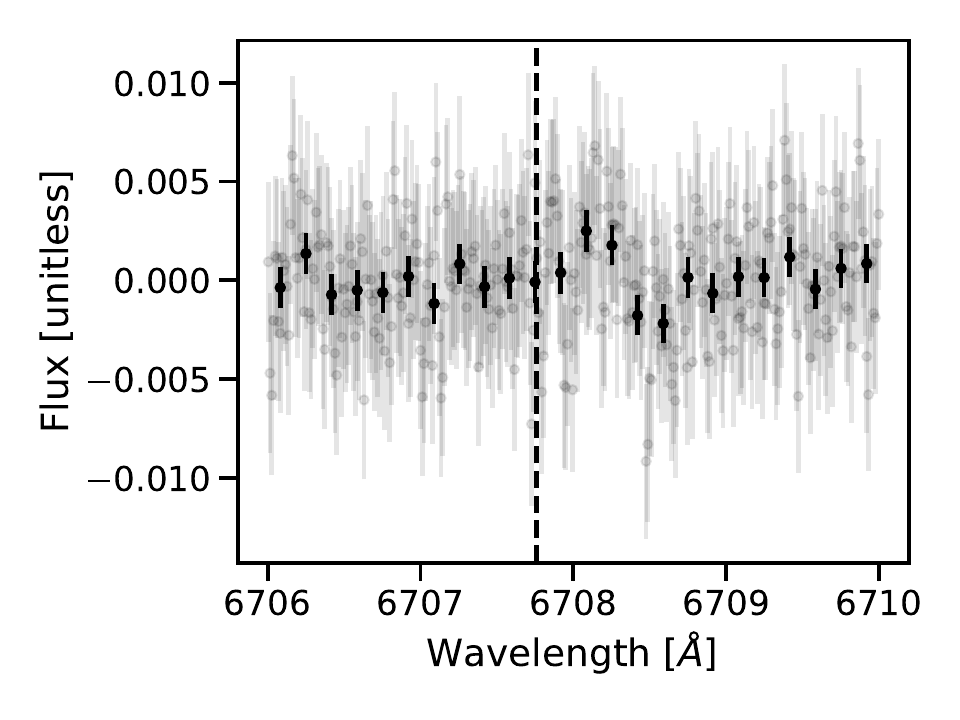}
    \caption{Transmission spectrum of potassium (top) and lithium (bottom). The black dashed line marks the line position; the solid red line shows the no-atmosphere POLD model (top panel only) to illustrate the deformation scale; the red dashed line shows the continuum (bottom panel only). Gray data points indicate the full transmission spectrum, with black bins every 15 pixels. Potassium is contaminated by POLD effects, with no obvious atmospheric detection. Lithium shows no detection, even though it is not contaminated by POLD effects, as this species does not absorb in the stellar spectrum.}
    \label{fig:resolved_lines}
\end{figure}

\begin{figure}
    \centering
    \includegraphics[width=\linewidth]{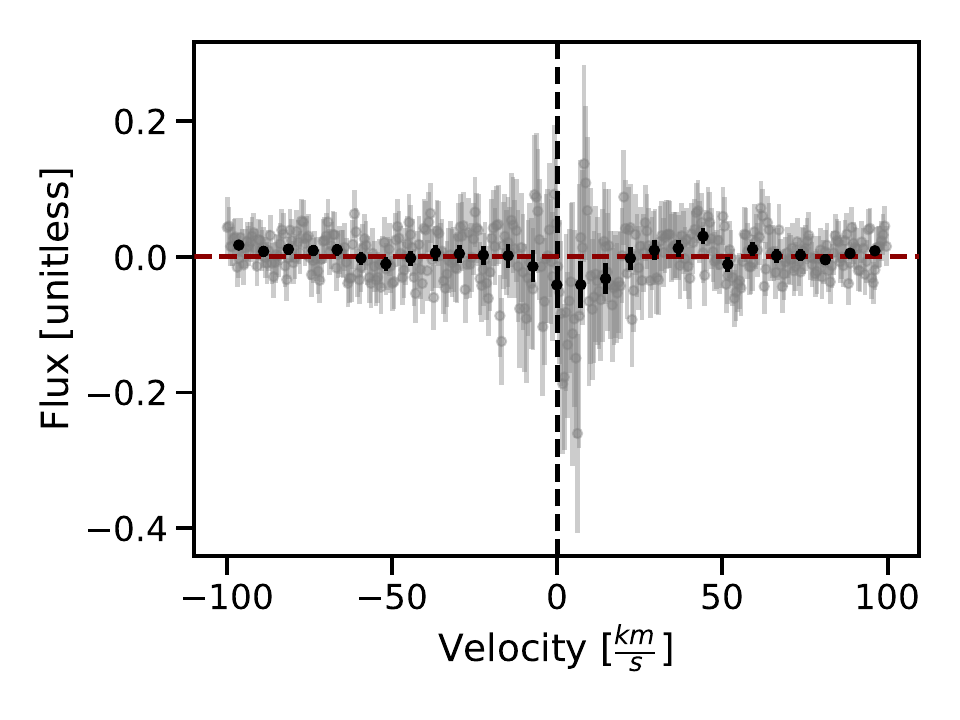}
    \includegraphics[width=\linewidth]{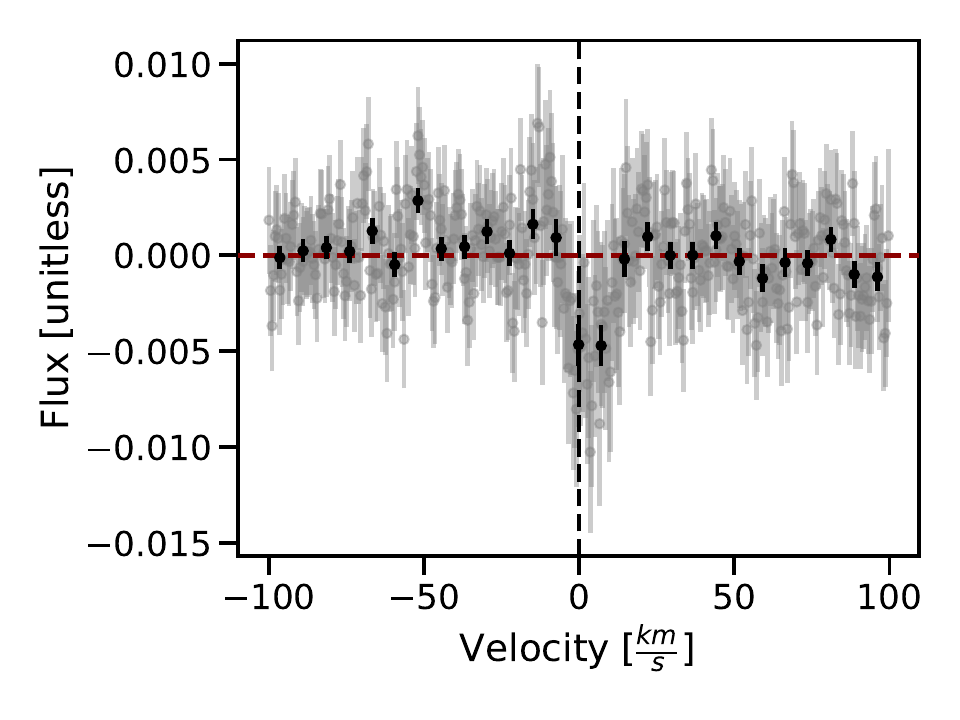}
    \includegraphics[width=\linewidth]{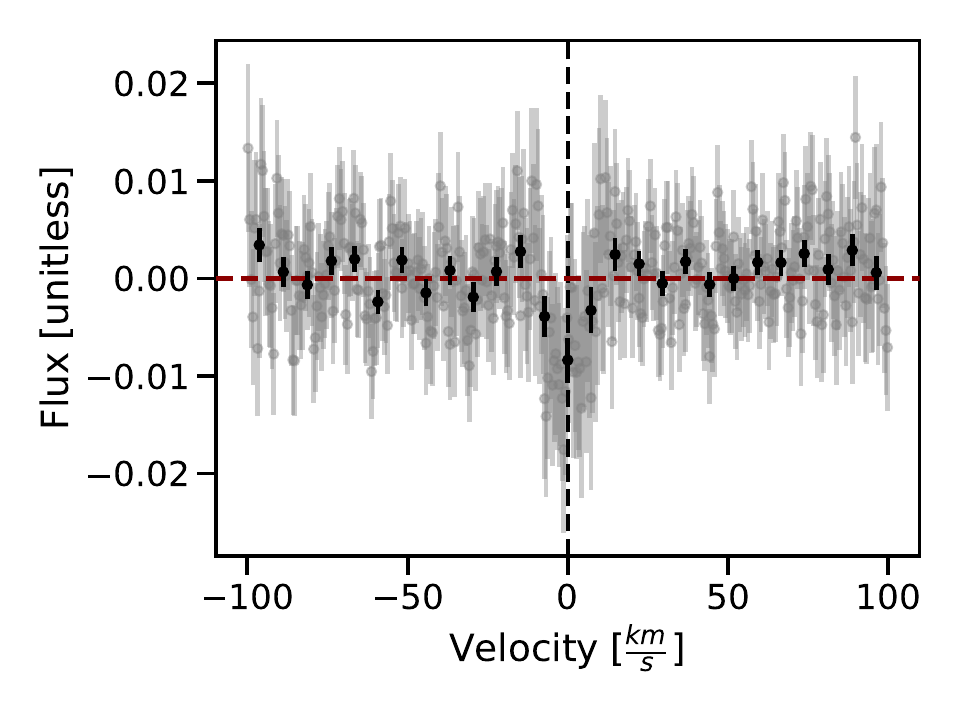}
    \caption{Velocity folded transmission spectrum of calcium (top), magnesium (middle) and manganese (bottom). The black dashed line is the 0 velocity, with the red dashed line being the continuum. Gray data points full transmission spectrum, with black bins by 15x pixels.  Strong POLDs contamination detected with no obvious signal.}
    \label{fig:velocity-folded-lines}
\end{figure}

We also looked at other lines besides \Na\,and \Ha. We followed the line-list by \cite{borsa2021}, and we velocity-folded each species together if there are multiple lines. In \autoref{fig:resolved_lines} we show single lines, in the wavelength-space, while in \autoref{fig:velocity-folded-lines} we show the velocity folded lines. No atmospheric feature is visible, only contamination by POLDs distortions (except for Li). The upper limits are reported in \autoref{tab:upper-limits-rest}, following the calculation described in \autoref{subsec:tsextraction}. As these species were not POLD corrected, we offset the line position by 1$\AA$ (wavelength space) or 20 kms$^{-1}$ (velocity space) from the laboratory position.

For potassium, we also calculate a no-atmosphere POLD distortion model with EvE, showing that the observed signal can be fully explained by POLDs. However, we note that we cannot exclude atmospheric presence without a full atmosphere model, similar to the sodium case in HD 209458 b \citep{dethier2023, carteret2024}. Furthermore, as we are sensitive only to a single line due to telluric contamination, we cannot exclude scenarios with a high ratio of amplitudes between the doublet lines.

\begin{table}[]
    \centering
    \caption{Upper limits for additional lines that were not corrected for POLDs.}
    \begin{tabular}{lr}
    \toprule
    Line & 3$\sigma$ upper limit \\
    \midrule
    $\text{K-7699} \AA $ & $ 0.65\%$ $ (1.18\,R_p)$ \\
    $\text{Li} $ & $ 0.66\%$ $ (1.18\,R_p)$ \\
    \midrule
    $\text{Mn} $ & $ 0.86\%$ $ (1.23\,R_p)$ \\
    $\text{Mg} $ & $ 0.51\%$ $ (1.15\,R_p)$ \\
    $\text{Ca} $ & $ 10.12\%$ $ (2.68\,R_p)$ \\
    \bottomrule    
    \end{tabular}
    \begin{minipage}{\linewidth}
    \vspace{0.1cm}
    \small Notes: The upper limit was calculated in the same way, but the line center was offset by 20 kms$^{-1}$ or 1 $\AA$. In parentheses is the corresponding height in planetary radii.
    \end{minipage}
    \label{tab:upper-limits-rest}
\end{table}

\section{Additional figures for analysis of the \Ha\,feature}
\label{subsec:halpha-feature}
\begin{figure}
    \centering
    \includegraphics[width=\linewidth]{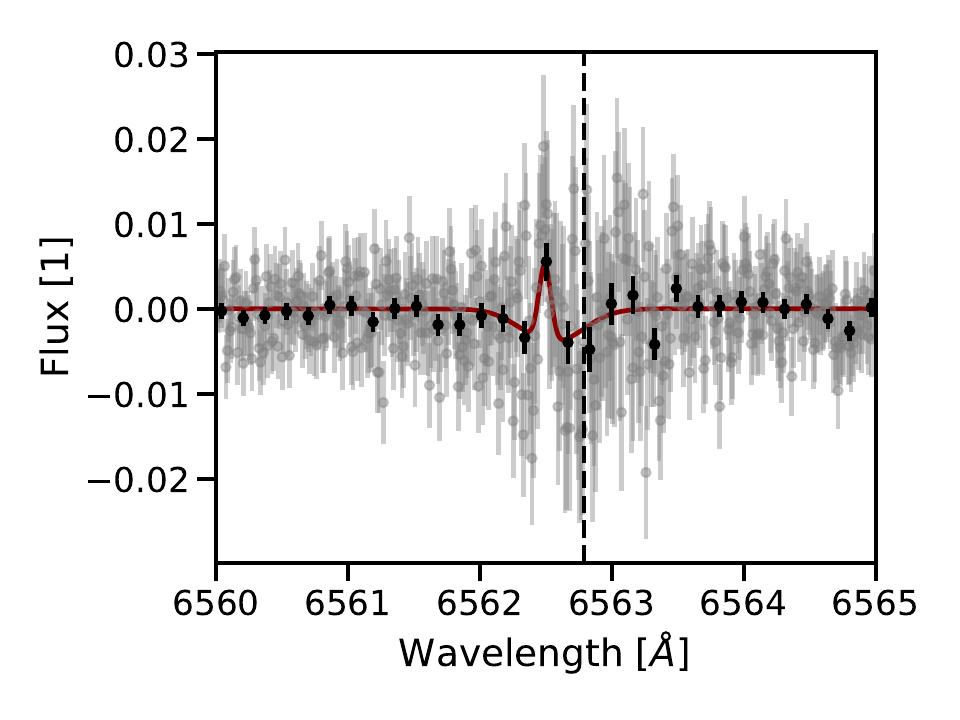}
    \caption{Fit of the double Gaussian to the RM+CLV corrected transmission spectra at the region of \Ha. }
    \label{fig:double-gaussian-fit}
\end{figure}

In this section, we include several additional figures for the \Ha\,feature discussion. In \autoref{fig:double-gaussian-fit}, we provide the best fit of double Gaussian model to the data, which provides a 1.6$\sigma$ significance to the amplitude. 

\begin{figure}
    \centering
    \includegraphics[width=\linewidth]{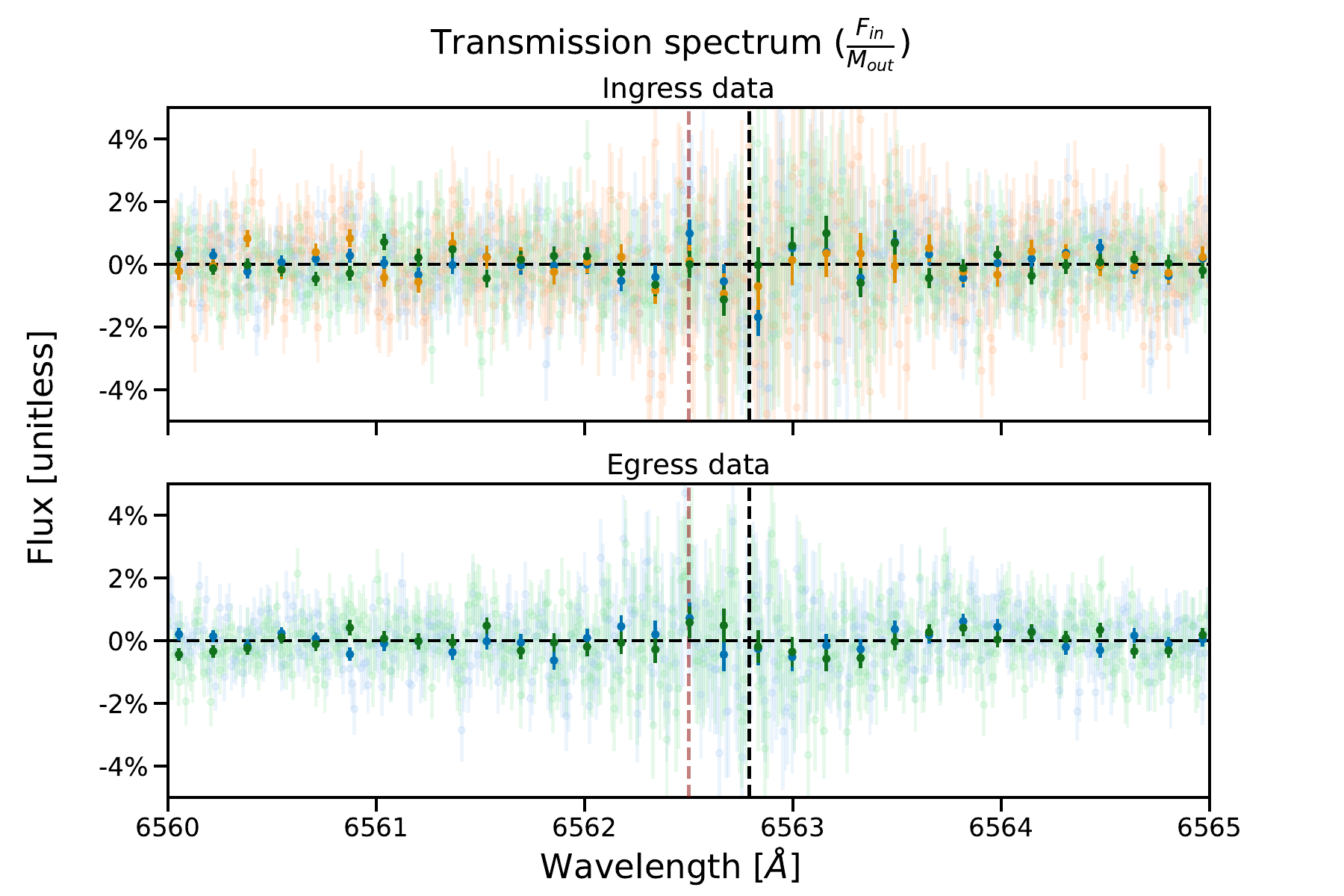}
    \caption{Transmission spectra in the region of \Ha\,per night as calculated for the “Ingress” data (top) and “Egress” data (bottom). Nights are color coded as blue, orange, and green for night \#1, \#2 and \#3 respectively. The vertical black dashed line corresponds to the laboratory wavelength of \Ha, with the red dashed line positioned at 6562.5$\AA$, the position of the observed absorption feature. The black dashed horizontal line is the continuum level. Note that the “Egress” data for night \#2 does not exist due to the partial transit observations.}
    \label{fig:transmission-per-night-per-phase}
\end{figure}

To further test the significance and repeatability of this feature, we divided our dataset in “Ingress” data ($\phi<0$; $\phi$ being the phase) and “Egress” data ($\phi>0$). We calculated an average transmission spectrum for the “Ingress” in-transit spectra and “Egress” in-transit spectra per night, which we show in \autoref{fig:transmission-per-night-per-phase}. Here, we show that the feature seems to be consistent (although a bit clearer in “Ingress” data), within our errorbars, and appear in all three nights consistently.

\begin{figure}
    \centering
    \includegraphics[width=\linewidth]{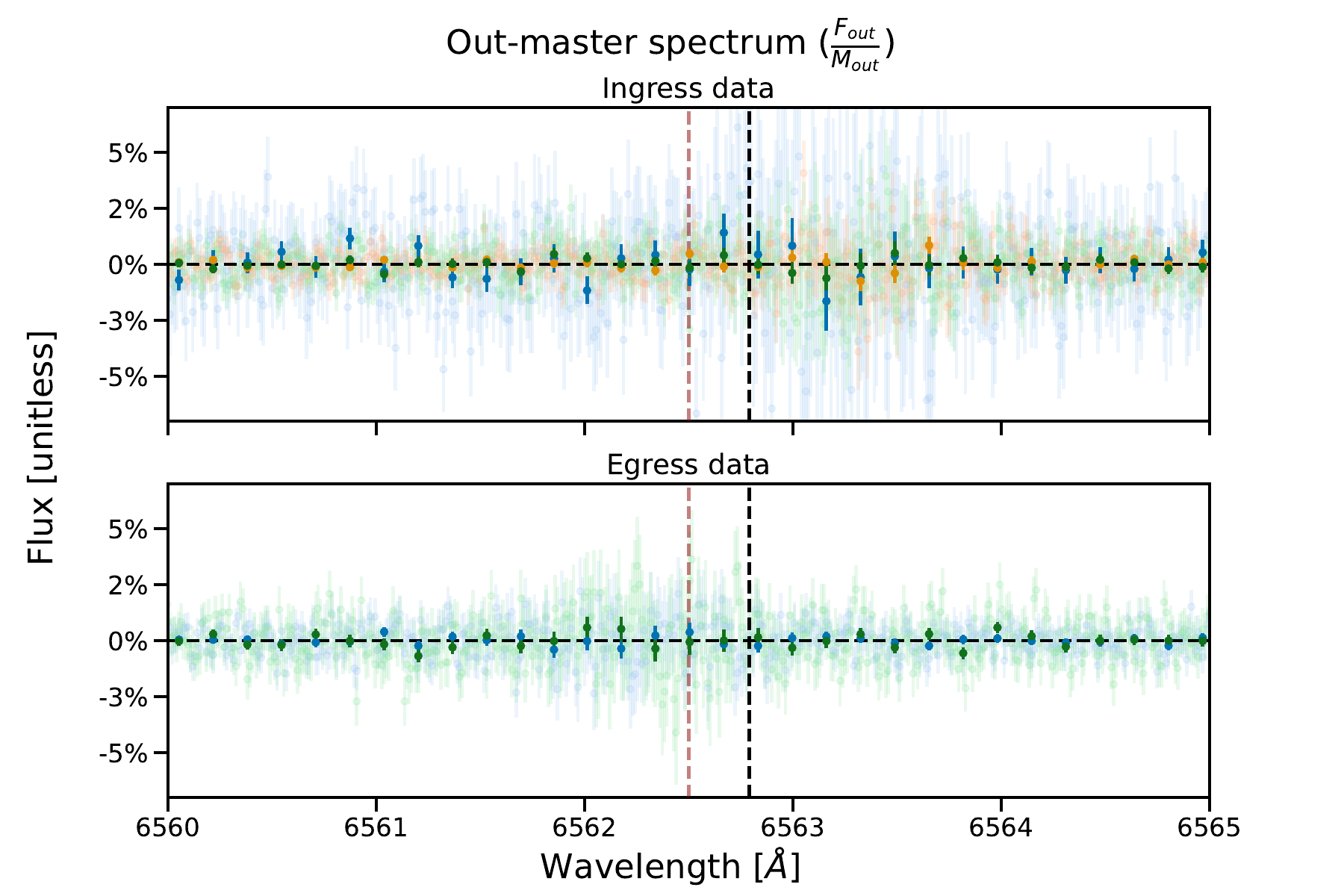}
    \caption{Out of transit spectra in the region of \Ha\,per night as calculated for the “Ingress” data (top) and “Egress” data (bottom). Nights are color coded as blue, orange, and green for night \#1, \#2 and \#3 respectively. The vertical black dashed line corresponds to the laboratory wavelength of \Ha, with the red dashed line positioned at 6562.5$\AA$, the position of the observed absorption feature. The black dashed horizontal line is the continuum level. Note the “Ingress” data for night \#1 consists of 3 spectra, which decreases the observed S/N.}
    \label{fig:out-per-night-per-phase}
\end{figure}

Finally, to test a possible systematic noise contamination by the star, we also calculate a “fake” transmission spectrum made out of out-of-transit data only (\autoref{fig:out-per-night-per-phase}), in the formula $\frac{F_{out}}{M_{out}}$. These spectra should always be flat, as long as our master-out spectrum is representative of the stellar spectra during the entire observation. This seems to be consistent well within the $1\sigma$ errorbars, showing the master-out stability during each night.

\end{appendix}

\end{document}